\documentclass[12pt,preprint]{aastex}
\usepackage{natbib}

\newcommand{\fluxunits}{\mbox{erg cm$^{-2}$ s$^{-1}$}}

\newcommand{\heiiw}{\mbox{\ion{He}{2} $\lambda$1640}}

\newcommand{\oiiw}{\mbox{[\ion{O}{2}] $\lambda$3727}}

\newcommand{\oiiiaw}{\mbox{[\ion{O}{3}] $\lambda$4959}}
\newcommand{\oiiibw}{\mbox{[\ion{O}{3}] $\lambda$5007}}
\newcommand{\oiiiw}{\mbox{[\ion{O}{3}] $\lambda \lambda$4959,5007}}

\newcommand{\lya}{\mbox{Ly$\alpha$}}

\newcommand{\hb}{\mbox{H$\beta$}}

\newcommand{\wrst}{\mbox{$W_{\lambda}^{\mbox{\tiny rest}}$}}
\newcommand{\wobs}{\mbox{$W_{\lambda}^{\mbox{\tiny obs}}$}}

\newcommand{\simlt}{\, \raisebox{-.2ex}{$\stackrel{<}{\mbox{\tiny $\sim$}}$} \,}

\newcommand{\bootes}{Bo\"{o}tes}

\newcommand{\fshrt}{\mbox{$f_{\tiny \nu}^{\mbox{\tiny short}}$}}
\newcommand{\flng}{\mbox{$f_{\tiny \nu}^{\mbox{\tiny long}}$}}
\newcommand{\ztit}{\mbox{$z \approx 4.5$}}
\newcommand{\vmax}{\mbox{$V_{\mbox{\tiny max}}$}}

\newcommand{\ntry}{\mbox{$\eta_{\mbox{\tiny try}}$}}
\newcommand{\flya}{\mbox{$f_{\mbox{\tiny \lya}}$}}
\newcommand{\fplya}{\mbox{$f^{\prime}_{\mbox{\tiny \lya}}$}}
\newcommand{\pdtct}{\mbox{$p_{\mbox{\tiny detect}}$}}

\newcommand{\wrsti}{\mbox{$W_{\lambda,i}^{\mbox{\tiny rest}}$}}
\newcommand{\nspec}{\mbox{$N_{\mbox{\tiny spec}}$}}
\newcommand{\ncand}{\mbox{$N_{\mbox{\tiny cand}}$}}
\newcommand{\fnb}{\mbox{$f_{\nu}^{\mbox{\tiny NB}}$}}
\newcommand{\fr}{\mbox{$f_{\nu}^{\mbox{\tiny R}}$}}

\begin{document}



\title{A Luminosity Function
of \lya-Emitting Galaxies at \ztit\altaffilmark{1,2}}

\author{
Steve Dawson\altaffilmark{3},
James E. Rhoads\altaffilmark{4},
Sangeeta Malhotra\altaffilmark{4},
Daniel Stern\altaffilmark{5},
JunXian Wang\altaffilmark{6},
Arjun Dey\altaffilmark{7},
Hyron Spinrad\altaffilmark{3},
Buell T. Jannuzi\altaffilmark{7}
}

\altaffiltext{1}{
Based in part on observations made at the W.M. Keck
Observatory, which is operated as a scientific
partnership among the California Institute of
Technology, the University of California, and the
National Aeronautics and Space Administration.  The
Observatory was made possible by the generous financial
support of the W.M. Keck Foundation.}

\altaffiltext{2}{Our data were 
obtained using community access telescope time made available
under the National Science Foundation's Telescope System 
Instrumentation Program (TSIP), awarded by the National Optical Astronomy 
Observatory.}

\altaffiltext{3}{
Department of Astronomy, University of California at
Berkeley, Mail Code 3411, Berkeley, CA 94720 USA;
sdawson@astro.berkeley.edu, spinrad@astro.berkeley.edu}

\altaffiltext{4}{
Arizona State University, Department of Physics and Astronomy
Arizona State University, P.O. Box 871504
Tempe, Arizona 85287-1504; Sangeeta.Malhotra@asu.edu, 
James.Rhoads@asu.edu}

\altaffiltext{5}{
Jet Propulsion Laboratory, California Institute of
Technology, Mail Stop 169--527, Pasadena, CA 91109 USA;
stern@zwolfkinder.jpl.nasa.gov.}

\altaffiltext{6}{
Center for Astrophysics, University of Science and
Technology of China, Hefei, Anhui 230026, China; jxw@ustc.edu.cn}

\altaffiltext{7}{
KPNO/NOAO, 950 N. Cherry Ave., P.O. Box 26732, Tucson,
AZ 85726 USA; dey@noao.edu, jannuzi@noao.edu}


\begin{abstract}
We present a catalog of 59 $z \approx 4.5$
\lya-emitting galaxies spectroscopically confirmed in
a campaign of Keck/DEIMOS follow-up observations to
candidates selected in the Large Area Lyman Alpha
(LALA) narrow-band imaging survey. We targeted 97
candidates for spectroscopic follow-up; by accounting
for the variety of conditions under which we performed
spectroscopy, we estimate a selection reliability of
$\sim 76\%$.  Together with our previous sample of
Keck/LRIS confirmations, the 59 sources confirmed
herein bring the total catalog to 73 spectroscopically confirmed $z \approx
4.5$ \lya-emitting galaxies in the $\approx 0.7$
degrees$^2$ covered by the LALA imaging. As with the
Keck/LRIS sample, we find that a non-negligible fraction
of the confirmed \lya\ lines have rest-frame
equivalent widths (\wrst) which exceed the maximum
predicted for normal stellar populations:  17\% -- 31\% (93\% confidence)
of the detected galaxies show $\wrst > 190$ \AA,
and 12\% -- 27\% (90\% confidence) show $\wrst > 240$ \AA.
We construct a luminosity function of $z \approx 4.5$
\lya\ emission lines for comparison to \lya\ luminosity
functions spanning $3.1 < z < 6.6$.  We find no significant
evidence for \lya\ luminosity function evolution from $z\approx
3$ to $z\approx 6$.  This result supports the conclusion that the
intergalactic medium remains largely reionized from the local universe
out to $z \approx 6.5$. It is somewhat at odds with the pronounced
drop in the cosmic star formation rate density recently measured
between $z \sim 3$ and $z \sim 6$ in continuum-selected Lyman-break galaxies, 
and therefore potentially sheds light on the relationship between the two
populations.
\end{abstract}

\keywords{cosmology: observations --- early universe
--- galaxies: evolution --- galaxies: formation ---
galaxies: high-redshift}

\righthead{Dawson et al.}
\lefthead{\lya\ Luminosity Function at $z \approx 4.5$}


\section{Introduction}
\label{introduction}

Observational cosmology has recently
witnessed a tremendous increase in proficiency in
the identification of galaxies at the earliest cosmic
epochs. Thanks in large part to the availability
of large-format mosaic CCDs well-suited for
wide-field imaging and spectroscopic multiplexing, we
are now transitioning from exotic, single detections of
high-redshift galaxies \citep[e.g.,][]{dey98, weymann98,
ellis01, ajiki02, dawson02, hu02, cuby03, taniguchi03,
nagao04, rhoads04, stern05} to the assembly of statistically
robust samples spanning the earliest accessible
redshifts.  Robust samples of this kind are necessary
for understanding the systematics of selection
criteria, and of the spatial distribution of the
galaxies themselves. Deficiencies in such understanding are the main
source of uncertainty in inferred luminosity functions
and universal star formation rates, which in turn are
the keys to understanding the cosmic history of star
formation, galaxy assembly and evolution, and even the
early ionization history of the IGM
\citep[e.g.,][]{malhotra04,stern05}.

Searches for high-redshift galaxies typically follow
the by-now familiar strategy of targeting redshifted
\lya\ emission at increasing wavelengths with
narrow-band imaging in windows of low night-sky
emission \citep[e.g.,][]{cowie98, hu98, rhoads00,
kodaira03, maier03, hu04, taniguchi05}, or by
photometric selection in broad-band imaging of the
redshifted Lyman break \citep[e.g.,][]{steidel96,
madau96, lowenthal97, spinrad98, lehnert03, ando04, bouwens04,
dickinson04, ouchi04, stanway04a, stanway04b, yan04}. These
two techniques are complementary; \lya\ searches at
typical sensitivities can identify galaxies with
UV-continua too faint to be detected by the Lyman
break method, but such surveys only select
that fraction of galaxies with strong line emission.

The Large Area Lyman Alpha (LALA) survey
\citep{rhoads00} has recently identified in deep
narrow-band imaging a large sample of \lya-emitting 
galaxies at redshifts $z \approx 4.5$
\citep{malhotra02}, $z \approx 5.7$ \citep{rhoads01,
rhoads03}, and $z \approx 6.5$ \citep{rhoads04}.  In
\citealt{dawson04} (Paper I), we reported on the
spectroscopic confirmation with the W. M. Keck Observatory's Low 
Resolution Imaging Spectrometer \citep[LRIS;][]{oke95} of 17
\lya-emitting galaxies selected in the LALA $z \approx
4.5$ survey. The resulting sample of confirmed \lya\   
emission lines showed large equivalent widths (median
$\wrst \approx 80$ \AA) but narrow velocity widths
(FWHM $\Delta v < 500$ km s$^{-1}$), indicating that
the \lya\ emission in these sources derives from
star formation, not from AGN activity. 
Models of star formation in the early universe predict 
that a small fraction of \lya-emitting galaxies at $z \approx 4.5$
may be nascent, metal-free objects
\citep[e.g.,][]{scannapieco03}, and indeed we found with
90\% confidence that 3 to 5 of the confirmed sources  
exceed the maximum \lya\ equivalent width predicted for
normal stellar populations.  However, we did not detect the
\heiiw\ emission expected to be characteristic of
primordial star formation.  Specifically, the \heiiw\ flux in a
composite of the 11 highest resolution spectra in the  
Keck/LRIS sample was formally consistent with zero, with
a 2$\sigma$ (3$\sigma$) upper limit of 13\% (20\%) of
the flux in the \lya\ line.  In other words,
though these galaxies may be young, they show no
evidence of being truly primitive, Population III
objects.

We have recently more than quadrupled our catalog of   
spectroscopically-confirmed \lya-emitting galaxies at
$z \approx 4.5$ with a spectroscopic campaign using the
W. M. Keck Observatory's DEep IMaging Multi-object Spectrograp
\citep[DEIMOS;][]{faber03}, targeting candidates selected from the
LALA survey. Together with the detections presented
in Paper I (and accounting for minor overlap in the
samples), the 59 \lya-emitters confirmed with
Keck/DEIMOS bring the total catalog to 73 spectroscopically
confirmed $z \approx 4.5$ \lya-emitting galaxies in the $\approx 0.7$
degrees$^2$ imaged by LALA. In this paper, we
utilize these additional confirmations to update the
results of Paper I, and to construct a luminosity
function of $z \approx 4.5$ \lya\ emission lines for
comparison to \lya\ luminosity functions spanning $3.1
< z < 6.6$. We describe our imaging and spectroscopic
observations in \S~\ref{observations}, and we summarize
the results of the spectroscopic campaign in
\S~\ref{results}. In \S~\ref{discussion}, we
investigate the distribution of the \lya\ lines in equivalent width,
we construct
\lya\ luminosity functions for our sample and
for several extant samples, and we discuss the
implications of the luminosity functions for 
the relationship between \lya-emitters
and Lyman-break galaxies (LBGs), and for the
history of reionization.
Throughout this paper we adopt a $\Lambda$-cosmology with
$\Omega_{\mbox{\tiny M}} = 0.3$ and $\Omega_\Lambda =
0.7$, and $H_0 = 70$ km s$^{-1}$ Mpc$^{-1}$ \citep{spergel03}.  At
$z=4.5$, such a universe is 1.3 Gyr old, the lookback
time is 90.2\% of the total age of the Universe, and an
angular size of 1\farcs0 corresponds to 6.61 comoving kpc.

\section{Observations}
\label{observations}

\subsection{Narrow-Band and Broad-Band Imaging}
\label{obs_imaging}

\begin{figure}
\centering
\epsscale{1.0}
\plotone{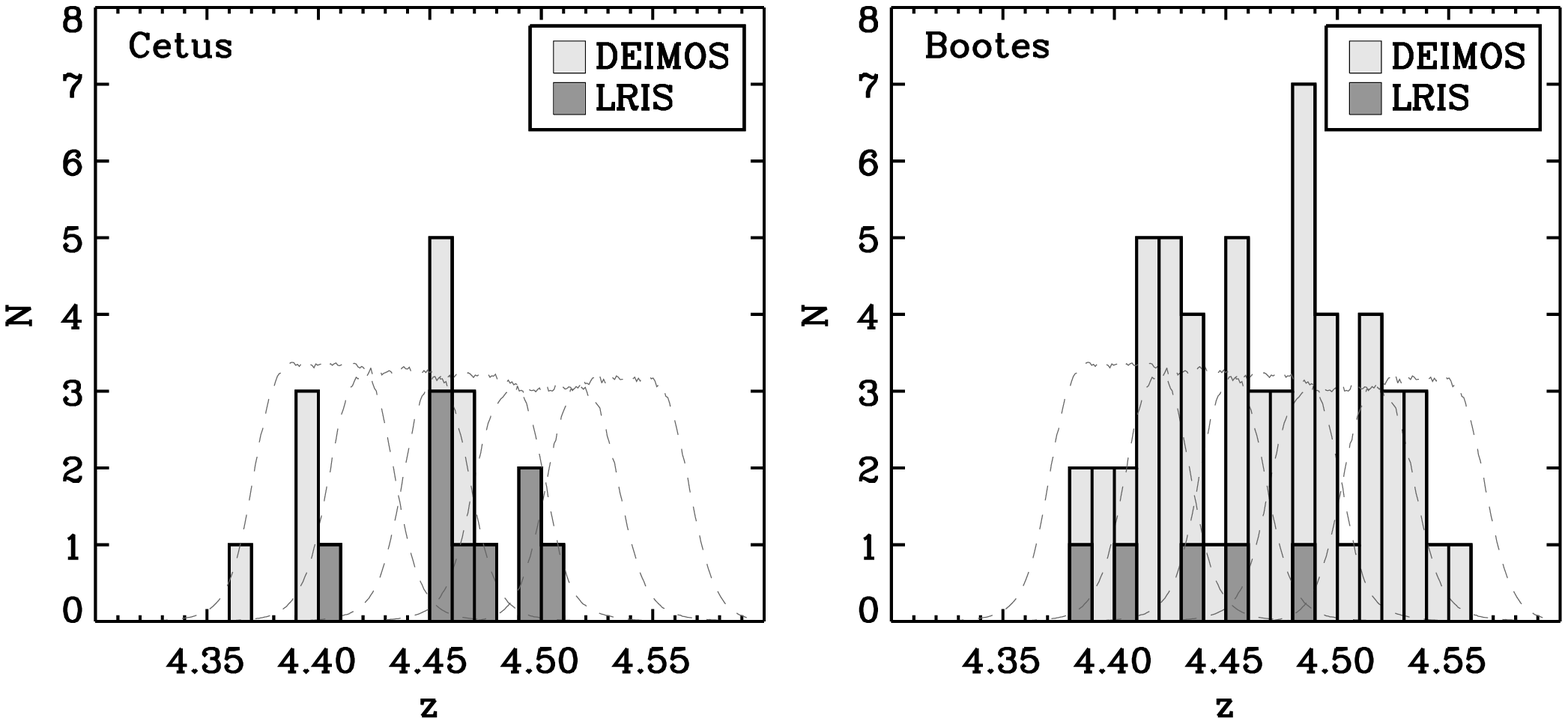}
\caption{Distribution of redshifts for
spectroscopically confirmed \lya\ emission lines in the
Cetus field (left; 02:05:20 $-$04:55, J2000.0) and in
the \bootes\ field (right; 14:25:57 $+$35:32, J2000.0).
The redshifts labeled ``DEIMOS" denote galaxies
confirmed with our campaign of Keck/DEIMOS
spectroscopy, described in this paper. The redshifts
labeled ``LRIS" denote galaxies confirmed with our
campaign of Keck/LRIS spectroscopy, described in Paper
I. The overlays are arbitrarily scaled transmission
curves for the five narrow-band filters employed in the
imaging component of this survey.
}
\label{z_hist}
\end{figure}

The LALA survey concentrates on two primary fields,
``\bootes" (14:25:57 $+$35:32; J2000.0) and ``Cetus''
(02:05:20 $-$04:55; J2000.0).  Each field is $36 \times
36$ arcminutes in size, corresponding to a single field
of the 8192 $\times$ 8192 pixel Mosaic CCD cameras on
the 4m Mayall Telescope at Kitt Peak National
Observatory and on the 4m Blanco Telescope at Cerro
Tololo Inter-American Observatory. The $z \approx 4.5$
search uses five overlapping narrow-band filters each
with full width at half maximum (FWHM) $\approx 80$
\AA\ (Figure \ref{z_hist}). The central wavelengths are
$6559$, $6611$, $6650$, $6692$, and $6730$ \AA, giving
a total redshift coverage of $4.37 < z < 4.57$ and a
survey volume of $7.4 \times 10^5$ comoving Mpc$^3$ per
field. In roughly 6 hours per filter per field, we
achieve 5$\sigma$ line detections in 2\farcs3 apertures
of $\approx 2 \times 10^{-17}$ \fluxunits.

The primary LALA survey fields were chosen to lie
within the NOAO Deep Wide-Field Survey
\citep[NDWFS;][]{jannuzi99}.  Thus, deep NDWFS
broad-band images are available in a custom $B_W$ filter
\citep[$\lambda_0 = 4135$ \AA, $\hbox{FWHM}=1278$ \AA;][Jannuzi 
et al., in preparation]{jannuzi99}
and in the Harris set Kron-Cousins $R$ and $I$, as well as
$J$, $H$, $K$, and $K_s$. The LALA \bootes\ field
benefits from additional deep $V$ and SDSS $z'$ filter
imaging. The imaging data reduction is described in
\citet{rhoads00}, and the candidate
selection is described in \citet{rhoads01} and \citet{malhotra02}. 
Briefly, candidates are selected based on a 5$\sigma$ detection
in a narrow-band filter, the flux density of which must
be twice the $R$-band flux density, and must exceed
the $R$-band flux density at the 4$\sigma$ confidence
level. To guard against foreground interlopers, we require
an observed equivalent width $\wobs > 80$ \AA, and a non-detection
in the $B_W$ band (at the $< 2\sigma$ level).

\subsection{Spectroscopic Observations}
\label{obs_spectroscopy}

Between 2003 March and 2004 May we obtained
spectroscopy of 97 $z \approx 4.5$ candidate
\lya-emitters with the DEep IMaging Multi-object
Spectrograph \citep[DEIMOS;][]{faber03}, a second-generation
camera on the Keck~II telescope with high
multiplexing capabilities and improved red sensitivity.
Each slitmask included approximately 15 candidate
\lya-emitters (mixed in with roughly 50 other
spectroscopic targets)  and was observed for 1.5 to
2.0~hrs in 0.5~hr increments.  Six slitmasks targeting
a total of 80 candidates were observed in the \bootes\
field; the airmass in these observations never exceeded
1.5.  One slitmask targeting 17 candidates was observed
in the Cetus field; the airmass for this slitmask was
constrained to less than 1.8. The seeing in all
observations ranged from 0\farcs5 to 1\farcs0.
We estimated the seeing by examining the alignment
stars observed during the direct-imaging phase of setting
up the slitmask; that is, the stars were imaged
through the BAL12 (clear) filter with the grating
angle set such that zero-order light fell on the detector.

All observations employed 1\farcs0 wide slitlets and
the 600ZD grating ($\lambda_{\rm blaze} = 7500$ \AA;
0.65 \AA\ pixel$^{-1}$ dispersion;  $\Delta
\lambda_{\rm FWHM} \approx 4.5$ \AA\
$\approx 200$ km s$^{-1}$)\footnote{We
measured the instrumental resolution by autocorrelating
one-dimensional extracted spectra of night-sky
emission lines.  The autocorrelation results in an
effective average line profile with a high
signal-to-noise ratio, which we fit with a Gaussian to
obtain the FWHM. We performed this test on $\sim 50$
night-sky spectra with the result $\Delta \lambda_{\rm
FWHM} = 4.47 \pm 0.03$ \AA.   The quoted uncertainty 
is the error in the mean, and does not include possible
systematic effects due to blended night sky lines.}.  
The wavelength range covered by a typical slitlet was
roughly 5000 \AA\ $\lesssim \lambda \lesssim 10000$ \AA.
The precise wavelength coverage depended somewhat on the location of the slitlet
on the slitmask, but never exceeded 4390 \AA\ at
the low extremum or 1.1 $\mu$m at the high extremum.
No order-blocking filter was used; since the targets were
primarily selected to have red colors, second order
light should not be of concern. Most nights suffered
from some cirrus;  relative flux calibration was
achieved from observations of standard stars from
\citet{massey90} observed during the same observing
run.  It should also be noted that the position angle
of an observation was set by the desire to maximize the
number of targets on a given slitmask, so observations
were generally not made at the parallactic angle.

We processed the two-dimensional data using the DEEP2
DEIMOS pipeline\footnote{See {\tt
http://astron.berkeley.edu/$\sim$cooper/deep/spec2d/}}.
We performed small (0\farcs5) dithers between exposures
on our initial observing run; to reduce these data, we
supplemented the DEEP2 DEIMOS pipeline with additional
home-grown routines. We extracted spectra with the
IRAF\footnote{IRAF is distributed by the National
Optical Astronomy Observatory, which is operated by the
Association of Universities for Research in Astronomy,
Inc., under cooperative agreement with the National
Science Foundation.} package \citep{tody93} using the
optimal extraction algorithm \citep{horne86}, following
standard slit spectroscopy procedures.

Prior experience with faint-object spectroscopy
dictates that a small but significant error in the
measured flux of faint continua may be introduced by
sky subtraction during the processing of the
two-dimensional spectra.  We investigated this
possibility with $\sim 10$ additional non-overlapping
extractions in source-free regions in each
two-dimensional spectrum, parallel to and along the
same trace as the extraction of the neighboring
\lya-emitting galaxy. We then fitted these blank-sky
spectra over the same region that we fitted for the
continuum redward and blueward of the emission line in
the object extraction. For 15 sources, these fits
yielded a tiny residual signal, which we interpreted as
a systematic error in the two-dimensional sky
subtraction and applied as a correction to quantities
derived from the object spectra.  The typical
correction was $\sim 0.04 \pm 0.02$ $\mu$Jy, but the
correction reached as high as $\sim 0.1 \pm 0.07$
$\mu$Jy in three cases.  Sky-subtraction residuals of this kind
generally resulted when a small spectroscopic slit
contained a bright serendipitous detection in addition
to the target, the combination of which made it
difficult to fit the sky background.

\section{Spectroscopic Results}
\label{results}

\begin{figure}
\centering
\epsscale{1.0}
\plotone{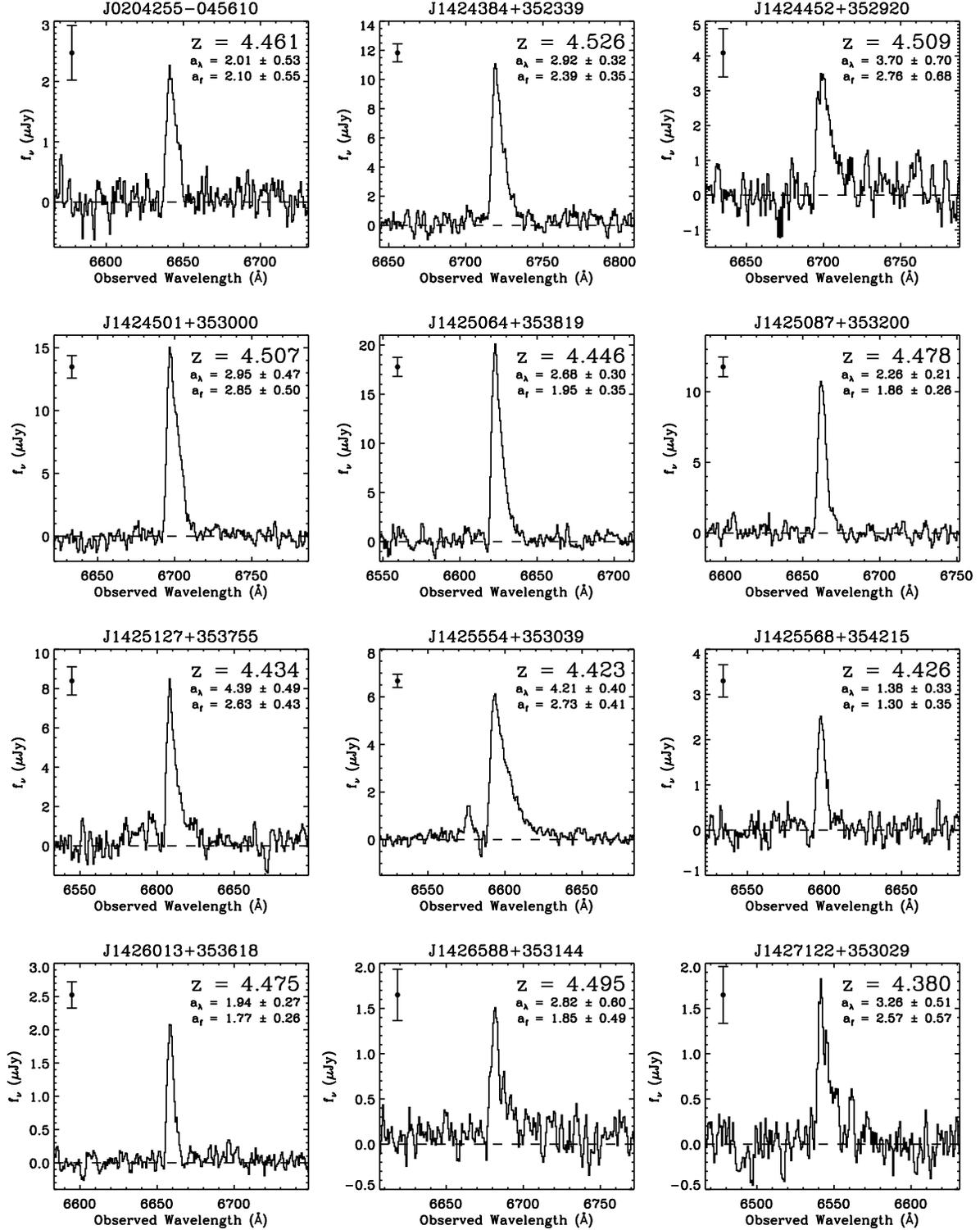}
\caption{\footnotesize Sample spectra from the set of 59 $z \approx
4.5$ \lya-emitting galaxies confirmed with
Keck/DEIMOS, with a wavelength range selected to
highlight the emission-line profile.  The measured
redshifts and asymmetry statistics
(\S~\ref{redshift_id}) are indicated in the upper right
of each panel.  The representative error bar ({\it
upper left}) is the median of the flux error in each
pixel over the wavelength range displayed.  The spectra
have been smoothed with a 3-pixel boxcar average.
}
\label{profiles}
\end{figure}

Out of 97 spectroscopic candidates, we achieved 73
detections, 59 of which constitute \lya\ confirmations
according to the criteria outlined below.  A histogram
of these confirmations appears in Figure~\ref{z_hist},
and a set of sample spectra are shown in
Figure~\ref{profiles}. The spectroscopic properties of
the \lya\ confirmations are summarized in
Table~\ref{table_spec_prop}. One detected galaxy lacks
an emission line but shows a large spectral
discontinuity identified as the onset of foreground
\lya-forest absorption at $z=4.462$\footnote{A sufficiently
bright Lyman break galaxy can be selected as a narrowband excess 
object when the narrow filter lies redward of the \lya\ forest,
so that neutral hydrogen absorption significantly reduces
the broad band flux without affecting the narrow band flux.}. 
Three of the
detections are identifiably low-redshift interlopers
(two resolved \oiiw\ doublets at $z \sim 0.8$; one
complex of \oiiiw\ and \hb\ at $z \sim 0.3$) which
survived the candidate selection thanks to their
unusually high equivalent widths (e.g., $\wobs > 2000$
\AA).  
In ten cases, we see a possible low
signal-to-noise ratio ($\lesssim 1$) emission line located
at the correct location in both wavelength space and
physical position to be associated with the
narrow-band selected target.  However, even if these
``detections" are real, they cannot be reliably
identified as either \lya\ or as low-redshift
interlopers.
If these 10 cases were in fact low signal-to-noise
ratio detections of \lya\ emission, 
then the ``success rate" of the
Keck/DEIMOS campaign would be 72\%, identical to that
of the Keck/LRIS sample described in Paper I (but also
subject to all the caveats listed therein). 
We do not include unconfirmed sources in any of the
ensuing discussion.

The remaining 24 targets were classified as
nondetections.  Five of these slitlets suffered from
some kind of instrument or reduction issue, e.g., the
target was dithered off the slitlet and so did not
reproduce across the individual integrations, or
irregularities in the machining of the slitmask
resulted in defects in the data processing.  Of the
final 19 nondetections, 13 targets were observed under
adverse conditions (e.g., variable cloud coverage,
and/or poor seeing) for which the general spectroscopic
yield was low.  Our failure to confirm these targets as
$z \approx 4.5$ \lya-emitters should not be taken to 
bear on the efficacy of candidate selection.

Six nondetections were observed under photometric conditions
with subarcsecond seeing for which the spectroscopic
yield was otherwise high.  However, subsequent
inspection of the imaging revealed that five of these
targets were suboptimal candidates for one of a variety
of reasons: two candidates sit on weak satellite trail
residuals; one candidate appears in an initial epoch of
imaging but not in subsequent epochs, suggesting that
it is a variable source or a spurious detection;  two
candidates are marginal or irregular detections in the
imaging.  This leaves just one otherwise viable
candidate \lya-emitter that was not confirmed in
spectroscopy, even though the conditions for
spectroscopy were favorable.  Since this source
(J1424398$+$353801) was a single-band detection in the
narrow-band imaging, it is possible that it represents a
spurious false-positive and not a genuine candidate.
Given the large number ($10^{6.5}$) of independent
resolution elements in the images, we expect about one
false-positive at the $5\sigma$-level per LALA field
per narrow-band filter, and this number could be larger
if the noise properties of the image are not precisely
Gaussian \citep[see][]{rhoads03}.

\begin{figure}
\centering
\epsscale{1.0}
\plotone{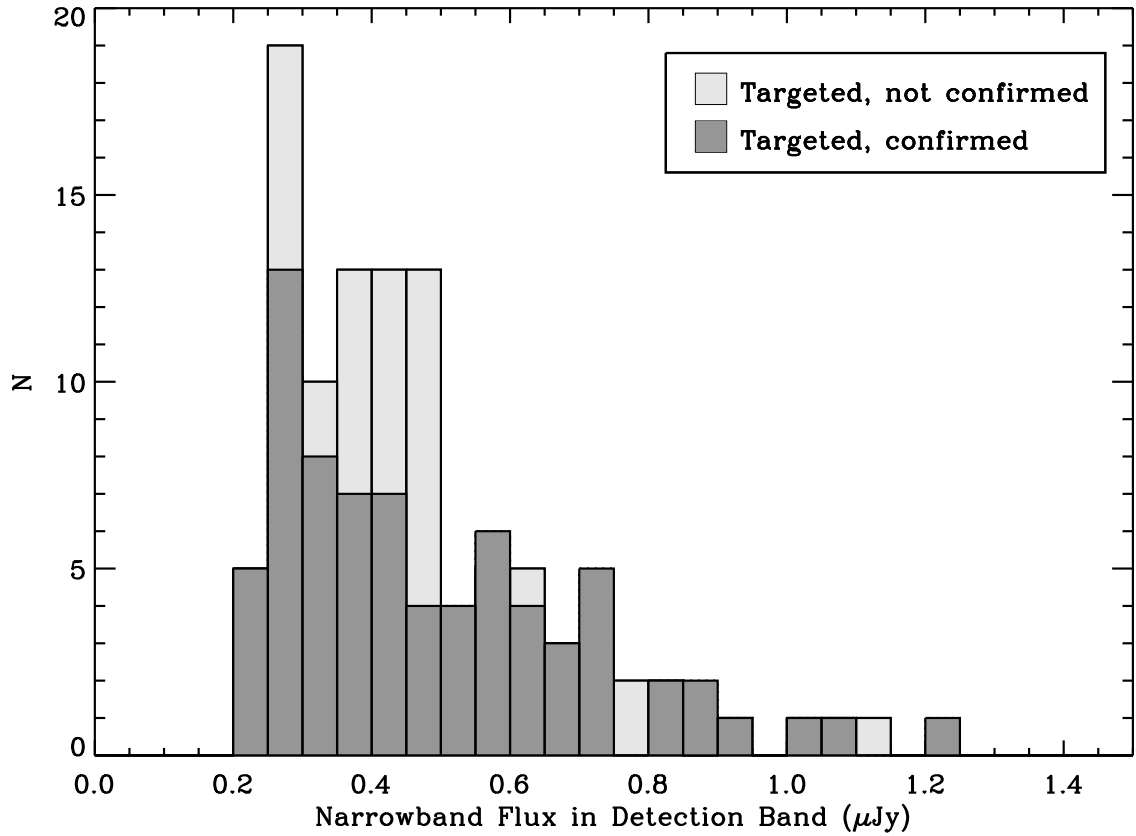}
\caption{Spectroscopic success rate as a function of
the flux in the narrow-band in which the candidate
was selected.  This plot combines the results of
the Keck/DEIMOS observations made for this paper and the
Keck/LRIS observations described in Paper I.}
\label{abs_success_frac}
\end{figure}

To estimate the reliability of our candidate
selection, we consider only the foregoing six
non-detections, the three low-redshift
interlopers, and the 10 low signal-to-noise
ratio detections as legitimate non-confirmations.
This admittedly rough scheme suggests a rate of
59 detections out of 78 viable candidates
observed spectroscopically under workable conditions,
for a final selection reliability of $\sim 76\%$.
The rate of spectroscopic confirmation is plotted 
as a function of narrow-band flux
in Figure~\ref{abs_success_frac}.

\subsection{Spectroscopic Sensitivity to \lya\ Emission}
\label{sensitivity_sec}

We now assess our spectroscopic sensitivity to line emission.
Each one-dimensional spectrum was created with a
variance-weighted optimal extraction \citep{horne86}
from the two-dimensional data. For each object, we
therefore have both the flux and the flux variance as a
function of wavelength.  We used the variance spectrum
to estimate the uncertainties in quantities derived
from the object spectra, e.g., we sum the variance spectrum
in quadrature over the wavelength range covered by the
observed line profile to estimate the uncertainty
in the measured line flux.
However, the variance may also be used to
measure the noise over wavelength ranges corresponding
to any \lya\ line we {\it might} have detected given
the redshift range permitted by our narrow-band imaging,
roughly $4.37 < z < 4.57$. Accordingly, for each object
we ranged over redshift and calculated the smallest
emission line flux detectable:
\begin{equation}
F_{\mbox{\tiny lim}}(z) = n{\mbox{\tiny sig}} \,
\delta_{\mbox{\tiny disp}} \left \{ \sum_{\lambda =
\lambda_1(z)}^{\lambda_2(z)} \sigma_\lambda^2 \right
\}^{\frac{1}{2}} \; ,
\end{equation}
where $n{\mbox{\tiny sig}}$ is the minimum
signal-to-noise ratio necessary for a detection (here taken
to be 3), $\delta_{\mbox{\tiny disp}}$ is the grating
dispersion (0.63 \AA\ pix$^{-1}$ for the Keck/DEIMOS 600ZD
grating), and $\sigma_\lambda$ is the flux error in
each pixel determined during the variance-weighted
one-dimensional extraction, in units of $f_{\lambda}$.
The limits $\lambda_1$ and $\lambda_2$ are defined by
\begin{equation}
\begin{array}{l}
\lambda_1(z) = (1+z)(1216 - \Delta \lambda / 2) \; , \\
\lambda_2(z) = (1+z)(1216 + \Delta \lambda / 2) \; , \\
\end{array}
\end{equation}
where $\Delta \lambda$ is the fiducial rest-frame full
width of the emission line (here taken to be 3 \AA).

\begin{figure}
\centering
\epsscale{1.0}
\plotone{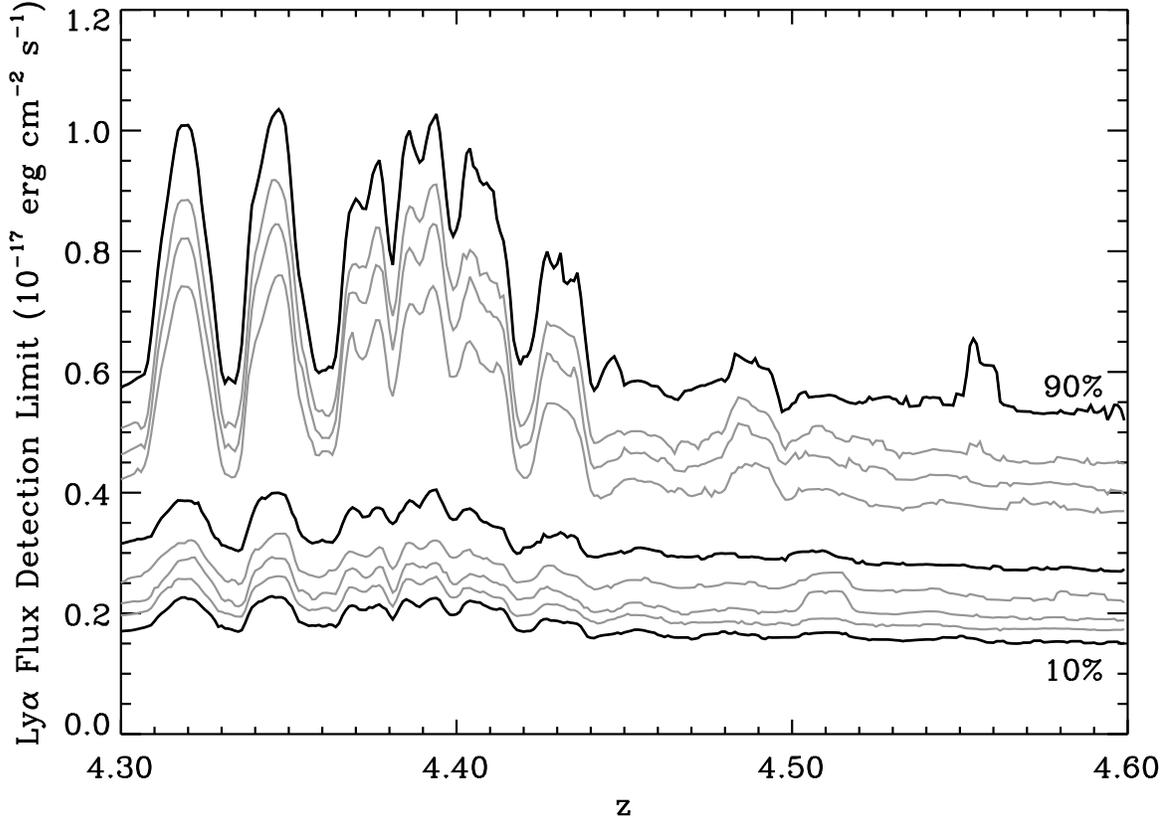}
\caption{Empirical, cumulative distribution of
spectroscopic sensitivity to \lya\ emission, as a
function of source redshift and \lya\ flux.  The
contours span 10\% to 90\% in 10\% steps.  The dark
lines denote the 10\%, 50\%, and 90\% contours.  The
distribution is plotted cumulatively so that it can be
interpreted as the probability that a putative \lya\
emission line of a given flux and redshift would have
been detected in our Keck/DEIMOS spectroscopic
campaign.
}
\label{spec_sens}
\end{figure}

We assembled the $F_{\mbox{\tiny lim}}(z)$ for each
object into a grid and then ranked the $F_{\mbox{\tiny
lim}}$ at each redshift, resulting in the cumulative
distribution of sensitivity to \lya\ emission line flux
shown in Figure \ref{spec_sens}. The distribution may
be interpreted as giving the probability that a
putative \lya\ emission line of a given flux and a
given redshift would have been detected in our
spectroscopic campaign.  Since we cover a comparatively
small redshift range centered essentially at the peak
of the detector throughput, the sensitivity
distribution is dominated entirely by night-sky
emission lines rather than by instrumental effects. And
since the original narrow-band survey was designed to
probe relatively noise-free windows in night-sky
emission, the spectroscopic sensitivity is fairly flat
over the redshift range of interest.  
In sum, the implied depth of our
spectroscopic survey is 50\% complete to $f(\lya) \sim
3 \times 10^{-18}$ \fluxunits, approximately 7 times deeper than the
narrow-band imaging.
Note that because
we derived this sensitivity function from the sample of
spectra themselves, it depends entirely on the details
governing the manner in which these spectra were
obtained and processed, and is therefore valid for this
survey only.

\subsection{Redshift Identification}
\label{redshift_id}

\begin{figure}
\centering
\epsscale{1.0}
\plotone{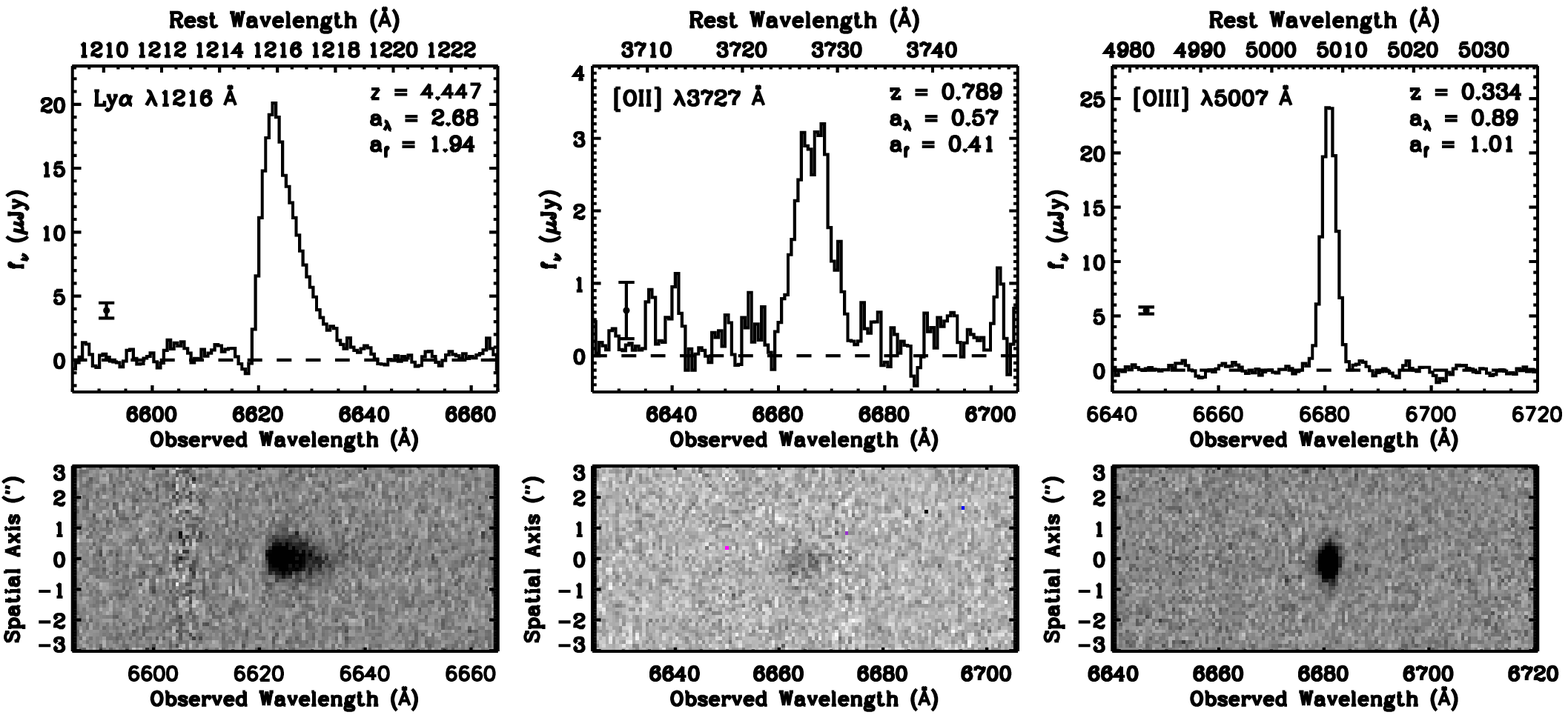}
\caption{Sample \lya\ emission line profile ({\it
left}) compared to two common low-redshift
interlopers: \oiiw\ ({\it center}) and \oiiibw\ ({\it
right}). The top figure in each case is the
one-dimensional extracted spectrum; the bottom figure
is a section of the two-dimensional data from which it
was extracted.  Note that we resolve the \oiiw\ doublet
with our Keck/DEIMOS spectroscopic setup, thereby
eliminating \oiiw\ as the main low-redshift interloper
in our survey.  The \oiiibw\ line can typically be
identified by neighboring \oiiiaw\ at one-third its
strength, or by neighboring \hb.
}
\label{oiifig}
\end{figure}

Of course, given the detection of an emission line, the
identification of that line as high-redshift \lya\ can
remain problematic. Thorough treatments of the pitfalls
of one-line redshift identifications are given
elsewhere \citep[e.g.,][]{stern99,stern00one,dawson01}.
In surveys of the present kind, the primary threat to
the proper interpretation of a solo emission line is the
potential for low-redshift, high-equivalent width
\oiiw\ to survive candidate selection, and then to be
misidentified as high-redshift \lya\ in later
spectroscopy. However, at $z=0.8$ (the redshift of an
\oiiw\ line mistaken for \lya\ at $z=4.5$), the
redshifted separation between the individual lines of
the \oiiw\ doublet (rest wavelengths 3726 \AA\ and 3729
\AA, respectively) is 5.4 \AA. The doublet is therefore
just resolved in our spectroscopy and serves to uniquely
flag \oiiw\ interlopers (Figure \ref{oiifig}); this is an
improvement afforded by Keck/DEIMOS over the
spectroscopy presented in Paper I. Less frequently,
high-equivalent width \oiiibw\ survives as an
interloper in our candidate selection.  However,
\oiiibw\ can typically be identified by neighboring
\oiiiaw\ at one-third its strength, or by neighboring
\hb.

\begin{figure}
\centering
\epsscale{1.0}
\plotone{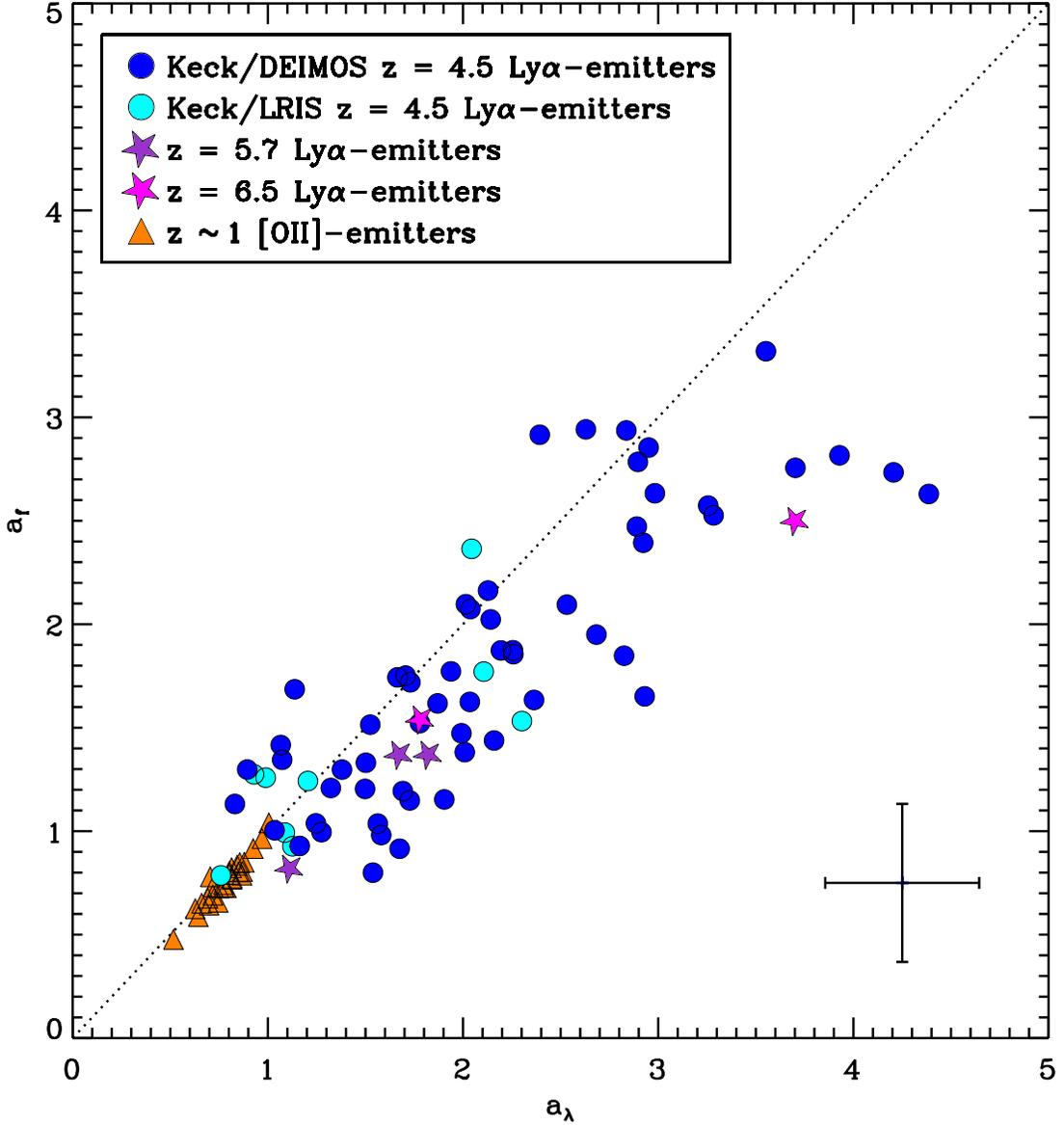}
\caption{\footnotesize Scatter plot comparing the flux-based
asymmetry statistic $a_f$ and the wavelength-based
asymmetry statistic $a_{\lambda}$ of known
high-redshift \lya-emitters to a sample of
\oiiw-emitters at $z \sim 1$, updated from Paper I.
The points labeled ``DEIMOS" denote galaxies confirmed
with our campaign of Keck/DEIMOS spectroscopy,
described in this paper. The points labeled ``LRIS"
denote galaxies confirmed with our campaign of
400${\ell}$/mm-grating Keck/LRIS spectroscopy,
described in Paper I. The three \lya-emitters at $z =
5.7$ are from \citet{rhoads03}, and the two
\lya-emitters at $z = 6.5$ are from \citet{rhoads04}
and \citet{stern05}. The 28
\oiiw-emitters at $z \sim 1$ were provided by the
DEEP2 team \cite[][A.\ Coil 2004, private
communication]{davis03}; their Keck/DEIMOS
1200$\ell$/mm-grating spectra were smoothed to the
Keck/LRIS 400$\ell$/mm-grating resolution by
convolution with a Gaussian kernel.  The representative
error bar ({\it lower right}) is the median of the
errors on the individual $a_f$ and $a_{\lambda}$ for
the combined Keck/LRIS and Keck/DEIMOS sample.
}
\label{asym_figure}
\end{figure}

Beyond merely eliminating plausible low-redshift
interlopers, we may identify \lya\ emission by its
characteristically asymmetric morphology, or by the
presence of a continuum break if the
continuum is sufficiently well-detected. Each of our confirmed
\lya\ detections demonstrates the asymmetric emission
line profile characteristic of the line, where neutral
hydrogen outflowing from an actively star-forming
galaxy imposes a sharp blue cutoff and broad red wing
\citep[e.g.,][]{dey98, stern99, manning00, dawson02, rhoads03,
hu04, stern05, taniguchi05}. In Figure
\ref{asym_figure}, we present a scatter plot of the
flux-based asymmetry statistic: \begin{equation} a_f =
\frac{\int_{\lambda_p}^{\lambda_{\mbox{\tiny 10},r}}
f_{\lambda} \, \mbox{d}
\lambda}{\int_{\lambda_{\mbox{\tiny 10},b}}^{\lambda_p}
f_{\lambda} \, \mbox{d} \lambda} \; , \end{equation}
versus the wavelength-based asymmetry statistic:
\begin{equation} a_{\lambda} =
\frac{(\lambda_{\mbox{\tiny 10},r} -
\lambda_p)}{(\lambda_p - \lambda_{\mbox{\tiny 10},b})}
\; , \end{equation} for our sample, where $\lambda_p$
is the wavelength of the peak of the emission line, and
$\lambda_{\mbox{\tiny 10},b}$ and $\lambda_{\mbox{\tiny
10},r}$ are the wavelengths at which the line flux
first exceeds 10\% of the peak on the blue side and on
the red side of the emission line, respectively
\citep[see][and Paper I]{rhoads03,rhoads04}\footnote{As
in Paper I, the error bars on $a_{\lambda}$ and $a_f$
were determined with Monte Carlo simulations in which
we modeled each emission line with the truncated
Gaussian profile described in \citet{hu04} and
\citet{rhoads04}, added random noise in each pixel
according to the photon counting errors, and measured
the widths $\sigma(a_{\lambda})$ and $\sigma(a_f)$ of
the resulting distributions of $a_{\lambda}$ and $a_f$
for the given line.  That is, for each $a_{\lambda,i}$,
the error $\delta a_{\lambda,i} =
\sigma(a_{\lambda,i})$, and similarly for each
$a_{f,i}$.}.  Each of the confirmed \lya\ emitters in
this sample satisfies $a_f > 1.0$ or $a_{\lambda} >
1.0$, and 52 out of 59 sources satisfy both. As 
we found for the lower resolution Keck/LRIS sample in 
Paper I, the present \lya\ sample, observed with higher spectral 
resolution using Keck/DEIMOS, is systematically segregated
from low-redshift \oiiw\ in $a_f$-$a_{\lambda}$ space.

As a final diagnostic, we note that in each of our
confirmed \lya-emitters for which the continuum is
sufficiently well-detected, the spectrum shows a
continuum decrement consistent with the onset of
absorption by the \lya\ forest at $\lambda_{\mbox{\tiny
rest}} = 1216$ \AA. The break amplitude is typically
characterized by $1 - \fshrt / \flng$, where we define
\fshrt\ as the variance-weighted flux density in a
1200 \AA\ window beginning 30 \AA\ below the emission
line; \flng\ is the same, but above the emission line.
In the 24 sources for which \flng\ is detected to
better than $2 \sigma$, all but two sources have $1 -
\fshrt / \flng > 0.5$, consistent with continuum break
amplitudes at $z=4.5$ in theoretical models
\citep[e.g.,][]{madau95, zhang97}, in the lower
resolution Keck/LRIS sample presented in Paper I, and
in other similar datasets \citep[see][and references
therein]{stern99}.

\section{Discussion}
\label{discussion}

Together with the observations presented in Paper I
(and accounting for minor overlap in the samples), the
59 \lya-emitters confirmed herein bring the total
catalog of spectroscopically confirmed $z \approx 4.5$ \lya-emitting 
galaxies to 73 objects in the $\approx 0.7$ degrees$^2$ surveyed
by the LALA imaging. We now update the characteristics of this
population as they were estimated in Paper I by
investigating the distribution of the total sample in  
equivalent width. We then
construct a $z \approx 4.5$ \lya\ luminosity function, carefully
accounting for survey incompleteness and for
spectroscopic sensitivity, and we compare the result to
\lya\ luminosity functions spanning $3.1 < z < 6.6$.

\subsection{The Equivalent Width Distribution}
\label{ew_update}

\begin{figure}
\centering
\epsscale{1.0}
\plotone{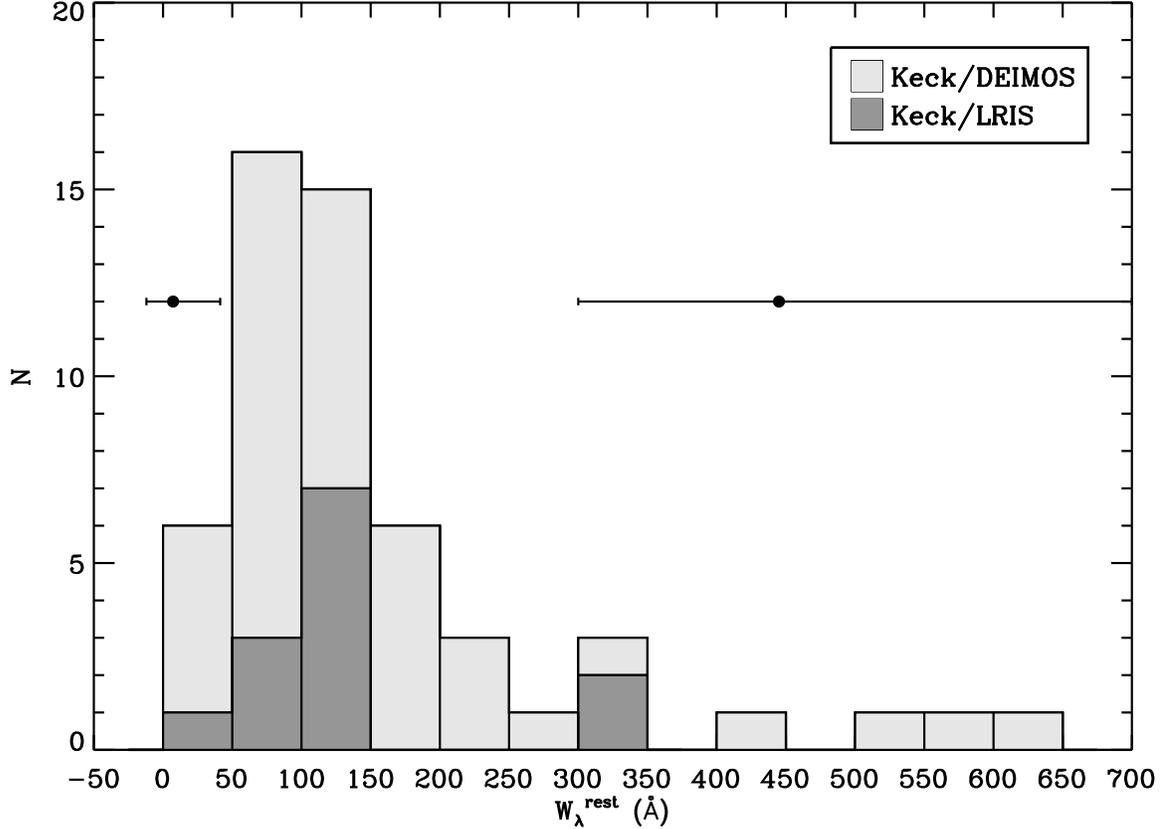}
\caption{Histogram of the spectroscopic rest-frame
equivalent widths for the $z=4.5$ population,
determined with $\wrst = (F_{\ell} / f_{\lambda,r}) /
(1+z)$, where $F_{\ell}$ is the flux in the emission
line and $f_{\lambda,r}$ is the measured red-side
continuum flux density. The sources labeled ``DEIMOS"
denote galaxies confirmed with our campaign of
Keck/DEIMOS spectroscopy, described in this paper. The
sources labeled ``LRIS" denote galaxies confirmed with
our campaign of Keck/LRIS spectroscopy, described in
Paper I. Representative error bars on the equivalent
widths are plotted at left and at right. Notably, the
highest equivalent widths are generally the least
certain, as they correspond to the faintest (and hence
least certain) continuum estimates.
}
\label{ew_hist}
\end{figure}

As in Paper I, we determine the rest-frame equivalent
widths directly from the spectra according to $\wrst = 
(F_{\ell} / f_{\lambda,r}) / (1+z)$, where $F_{\ell}$
is the flux in the emission line and $f_{\lambda,r}$ is
the measured red-side continuum flux density. The
resulting equivalent width distribution is plotted in 
Figure~\ref{ew_hist}, together with the equivalent
widths measured in Paper I.

Before interpreting this distribution,
one should be cautioned that the \wrst\ determination
is very sensitive to uncertainty in the measured
continuum. Since the continuum estimate enters into the
denominator of the expression for \wrst, the
characteristically small continuum values and their
large fractional uncertainties cause significant
scatter in the measurement, and the resulting error is
neither Gaussian nor symmetric about the measured
value.  Especially problematic is the fact that the
largest values of \wrst\ are also the least certain.
Detailed discussions of the uncertainties in measuring
\wrst\ in high-redshift \lya-emitters, along with the
complicating effects of dust content, gas kinematics,
and intergalactic absorption, are given in \citet{hu04}
and in Paper I.

With these caveats in mind, we rigorously treated the
error bars on the equivalent width estimates, and we
restricted the analysis to sources with red-side
continuum signal-to-noise ratios $\gtrsim 1$.  To determine
the equivalent width error bars, we first associated
each measured line flux $F_{\ell,i} \pm \delta
F_{\ell,i}$ with a Gaussian probability density
function (PDF) centered on $F_{\ell,i}$ with width
$\sigma = \delta F_{\ell,i}$; we proceeded similarly
for the measured continuum fluxes. We then generated a
grid of line flux versus continuum flux on which each
node has an associated equivalent width and is assigned
a weight given by the probability distribution on each
of its flux axes.  Next we collapsed the grid into a
histogram of equivalent widths, adding the weight from
each grid point to the appropriate equivalent width
bin.  The result is a non-Gaussian PDF $P_i (w)$ for
which $P_i (w) \, dw$ is the probability of observing
\wrsti\ in the interval $w < \wrsti\ < w + dw$. The
error bars $\delta w_{\mbox{\tiny +}}$ and $\delta
w_{\mbox{\tiny --}}$ are then $1 \sigma$ confidence
intervals determined by integrating over the
probability density functions $P_i(w)$.  They are
symmetric in probability density-space in the sense
that $\int_{w - \delta w_{\mbox{\tiny --}}}^w P_i(w')
\, dw' = \int^{w + \delta w_{\mbox{\tiny +}}}_w P_i(w')
\, dw'= 0.34$.

We find the resulting distribution to be broadly
consistent with the equivalent widths presented in
\citet{fujita03} for $z \sim 3.7$ and in \citet{hu04}
for $z \sim 5.7$.  While the majority of sources can
be understood as comparatively young (1 to 10 Myr)
galaxies with Salpeter initial mass functions (IMFs), a
non-negligible fraction exceeds the largest
rest-frame equivalent widths expected from such
stellar populations. \citet{malhotra02} use a Salpeter
initial mass function, an upper mass cutoff of 120
$M_{\sun}$, and a metallicity of 1/20th solar to find
maximum \lya\ equivalent widths of 300 \AA, 150 \AA,
and 100 \AA\ for stellar populations of ages $10^6$,
$10^7$, and $10^8$ years, respectively. Adopting a
correction factor of 0.64 as an upper limit to the
effect of IGM absorption on the measurement of \wrst\  
in spectroscopy effectively reduces these upper limits
to 190 \AA, 100 \AA, and 60 \AA (see discussion in
Paper I). Owing to the lower metallicity used in their
models, the pre-IGM-corrected values of
\citet{malhotra02} are slightly higher than the  
canonical limiting \lya\ rest-frame equivalent width
of 240 \AA\ given by \citet{charlot93}.

Using the ensemble of $P_i(w)$ described above, we find that
12\% -- 27\% (90\% confidence) of the galaxies in this sample show $\wrst
> 240$ \AA, and 17\% -- 31\% (93\% confidence) show
$\wrst > 190$ \AA.  Both results are
nearly identical to the values given in Paper I.
On the simplest interpretation, these 
galaxies are required to be very young (age $<
10^6$ years), or to have IMFs skewed in favor
of the production of massive stars.  
The possibility that AGNs in our sample are producing
stronger-than-expected \lya\ emission seems unlikely
due to the comparatively narrow velocity widths of the
\lya\ lines and to the absence of the
high-ionization state UV emission lines symptomatic of
AGN activity.  Moreover, deep ($\sim
170$ ks) {\it Chandra}/ACIS imaging of LALA $z \approx 4.5$
candidates in both \bootes\ \citep{malhotra03} and
in Cetus \citep{wang04} resulted in X-ray non-detections
to an average 3$\sigma$ limiting luminosity of $L_{2-8 \mbox{\tiny
keV}} < 2.8 \times 10^{42}$ erg s$^{-1}$.  This limit is roughly an order
of magnitude fainter than what is typically observed for even the heavily
obscured, Type II AGNs \citep[e.g.,][]{stern02q,norman02,dawson03}.
By comparing the upper limit on the typical X-ray to \lya\ luminosity ratio
for the \lya\ galaxy sample to the observed values of this ratio
for quasar and Seyfert galaxy samples, \citet{malhotra03} and 
\citet{wang04} conclude
that AGN account for $\la 5\%$ of the \lya\ galaxy sample.

\subsection{Empirical Cumulative Luminosity Function}
\label{cumlumfunc}

\begin{figure}[!t]
\centering
\epsscale{1.0}
\plotone{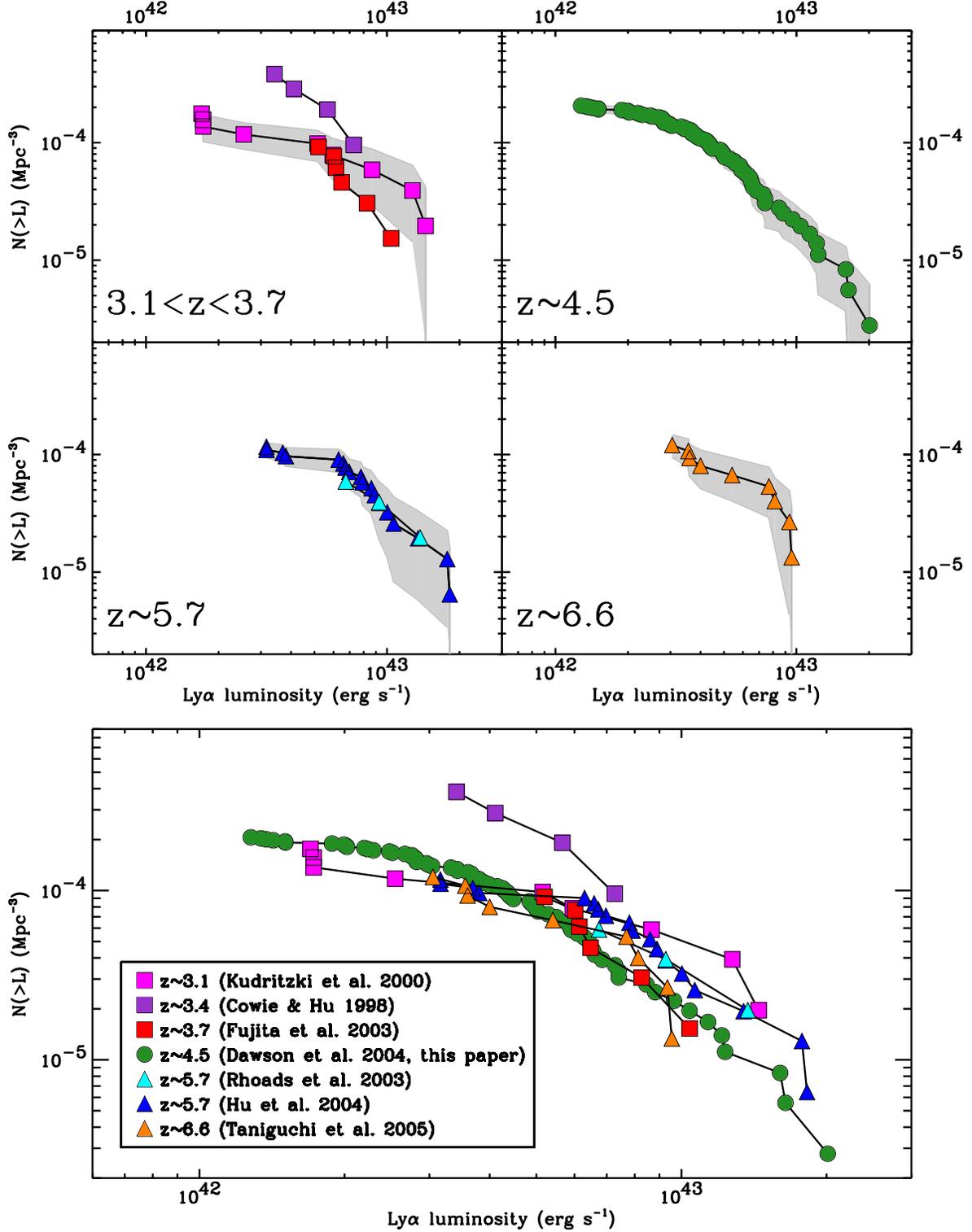}
\caption{\footnotesize Comparison of empirical, cumulative \lya\
luminosity functions computed with only minimal
completeness correction for several spectroscopic
surveys spanning $3.1 < z < 6.6$.  The cumulative luminosity function
gives for each \lya\ line luminosity $L(\lya)$ the
total number density of \lya\ lines brighter than
$L(\lya)$.  
The shaded regions represent 95\% confidence
intervals based on the Monte Carlo simulations
described in \S~\ref{cumlumfunc}.
Where more than one survey
is plotted, just the confidence intervals for the survey with the largest range
in line fluxes is depicted.
No strong evolution is evident over the
redshift range depicted. 
}
\label{lumfuncfig}
\end{figure}

In Figure~\ref{lumfuncfig}, we present an empirical
cumulative \lya\ line luminosity function computed
for our sample at $z \sim 4.5$ and compare this 
to luminosity functions computed
for several other samples spanning $3.1 < z < 6.6$. The
cumulative luminosity function gives for each \lya\ line luminosity
$L(\lya)$ the total number density of \lya\ lines
brighter than $L(\lya)$. The comparison samples are
drawn from spectroscopic follow-up of narrow-band
surveys with roughly comparable flux limits and
candidate-selection criteria (except where noted,
below).  
We do not include non-spectroscopic \lya-emitter
luminosity functions \citep[e.g.,][]{ouchi03}
among the comparison samples.
In each case, we converted the reported \lya\
line fluxes to line luminosities using a
$\Lambda$-cosmology with $\Omega_{\mbox{\tiny M}} =
0.3$ and $\Omega_\Lambda = 0.7$, and $H_0 = 70$ km
s$^{-1}$ Mpc$^{-1}$, and we made a minimal attempt to
account for incompleteness\footnote{\citet{hu04}
provide fluxes for their $z \sim 5.7$ sources as
measured in narrow-band imaging, rather than \lya\ line
fluxes as measured in spectroscopy.  As such, we adopt
the conversion given in \citet{stern05} to estimate
\flya\ from \fnb.}.
Specifically, the volume from which \lya-emitting
candidates were selected by their narrow-band excess is
simply defined by the solid angle covered by the
narrow-band imaging and the redshift range allowed  
by the narrow-band filter. However, the {\it effective}
volume surveyed by the spectroscopic follow-up is
smaller than the imaging survey volume by a factor of
$\nspec / \ncand$, where \ncand\ is the total number of
\lya-emitting candidates discovered in the imaging, 
and \nspec\ is the number of candidates actually
targeted for spectroscopy.
We estimated the uncertainties in the cumulative
luminosity functions with Monte Carlo simulations.
Assuming the
errors in the \lya\ line fluxes are Gaussian, we created
synthetic data sets by drawing randomly from the Gaussian
\lya\ flux PDFs for each object in each sample.
The result for each sample was then a distribution
of cumulative luminosity functions, which may be used to define upper and lower confidence
intervals.  Figure~\ref{lumfuncfig} depicts 95\%
confidence intervals; where more than one survey
is plotted, just the confidence intervals for the survey with the largest range
in line fluxes is depicted.

No strong evolution is readily evident in the cumulative \lya\
luminosity functions between $z \sim 3$ and $z \sim 6$. The only
significant scatter between luminosity functions occurs between the
various $z \sim 3$ surveys, and that scatter likely finds its
origin in differences in the manner in which the
experiments were performed. Foremost, the area surveyed
by the \citet{cowie98} effort is comparatively small:
just 25 arcmin$^2$ in each of two fields (HDF and SSA
22), as opposed to 300 arcmin$^2$ in
\citet{kudritzki00} and 132 arcmin$^2$ in
\citet{fujita03}.  \citet{cowie98} note that the number
counts in their HDF field appears to be 2.5 times
richer in narrow-band excess objects than their SSA 22
field, highlighting the susceptibility of small survey
areas to cosmic variance.  Separately, as noted by
\citet{hu04}, the \citet{fujita03} data may
comparatively under-represent the density of
\lya-emitters due to their more stringent equivalent
width criterion of $\wobs > 250$ \AA, as opposed to
$\wobs > 77$ \AA\ in \citet{cowie98} and effectively
$\wobs \gtrsim 100$ \AA\ in \citet{kudritzki00}.

\subsection{The {\rm $V / \vmax$} Estimate}
\label{difflumfunc}

\begin{figure}
\centering
\epsscale{1.0}
\plotone{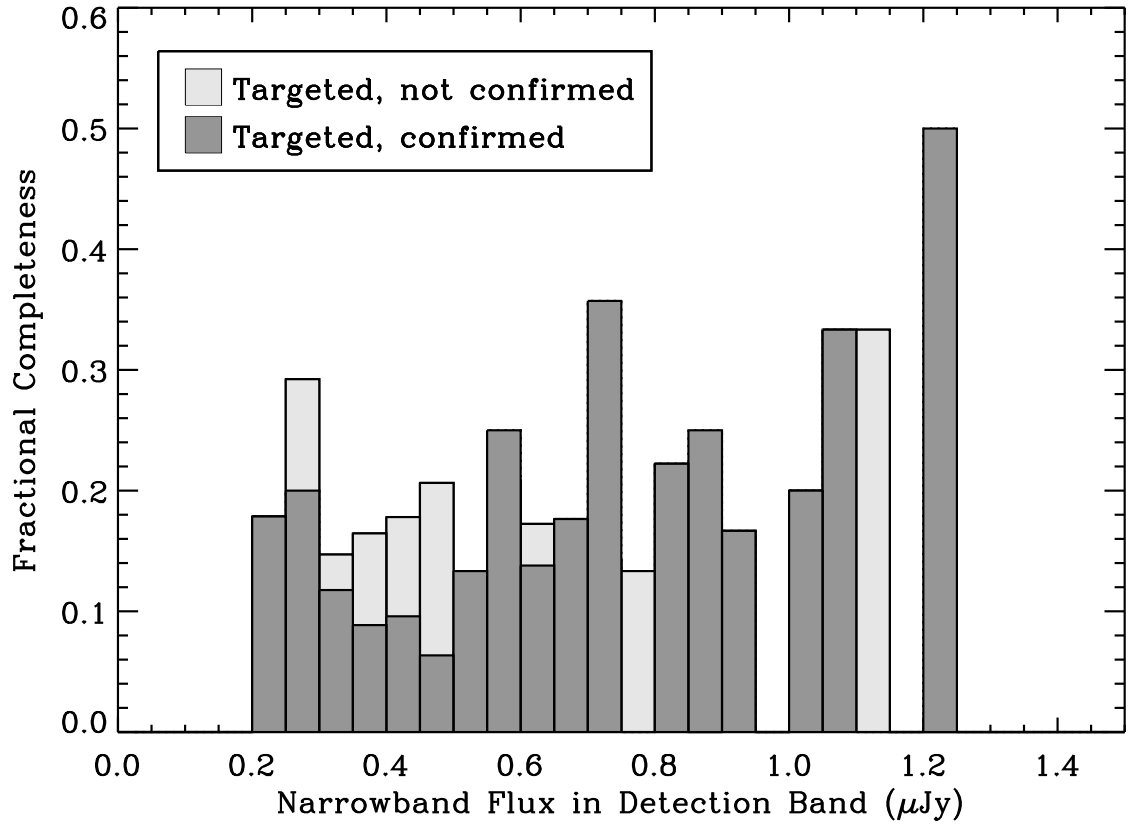}
\caption{Probability as a function of narrow-band flux
that a candidate \lya-emitter was targeted for
spectroscopy, 
divided into the fraction of targets that were confirmed
and the fraction of targets that were not.
}
\label{completeness}
\end{figure}

We now perform a more rigorous measurement of the $z\approx 4.5$
\lya\ luminosity function using a modified version of the $V / \vmax$ 
method \citep[e.g.,][]{hogg98, fan01}. 
For each galaxy, \vmax\ is the volume over which \lya\
of a given luminosity could be located and still be
detected by our survey; the luminosity function is then
the sum of the inverse volumes of all galaxies in the
given luminosity bins. Our modifications to the $V /
\vmax$ method account for incompleteness in two senses.
First, not every galaxy candidate identified in imaging
was targeted in follow-up spectroscopy.  Following
\citet{hogg98}, Figure~\ref{completeness} shows the
fraction of narrow-band selected candidate
\lya-emitters which were targeted for spectroscopy as
a function of flux in the band in which the candidate
was detected.  We label this {\it a priori}
completeness function \ntry; the candidate \lya\ flux
\flya\ can be roughly estimated from the flux in the
narrow-band \fnb\ with $\flya = w_n (\fnb - \fr)$, where
$w_n$ is the width of the narrow-band filter and \fr\
is the flux of the candidate in the $R$-band.

Second, even if a candidate \lya-emitter was selected
for spectroscopy, its inclusion in the luminosity
function depends on the detection and identification of
the \lya\ line.  Our spectroscopic sensitivity to \lya\
emission as a function of flux and redshift is shown in
Figure~\ref{spec_sens};  we label this function \pdtct.
As discussed in section \S~\ref{sensitivity_sec},
\pdtct\ can be interpreted as the probability that a
putative \lya\ emission line of a given flux and a
given redshift would have been detected in our
spectroscopic campaign.

In the presence of these selection effects, the
available volume for a galaxy with \lya\ emission of
flux \flya\ is
\begin{equation}
\vmax = \int_{z_1}^{z_2} \ntry ( \fplya ) \, \pdtct
(\fplya,z') \, \frac{d^2V_c}{d \Omega \, d z'} \Delta
\Omega \, dz' \; ,
\label{eq_vmax}
\end{equation}
where the comoving volume element in a solid angle $d
\Omega$ and redshift interval $dz$ is the familiar
\begin{equation}
\frac{d^2V_c}{ d \Omega \, dz} = \left ( \frac{c}{H_0}
\right )^3 \left \{ \int_0^z \frac{dz'}{E(z')} \right
\}^2 \frac{1}{E(z)} \; ,
\end{equation}
with
\begin{equation}
E(z) = \left \{ \Omega_{\mbox{\tiny M}} (1+z)^3 +
\Omega_{\mbox{\tiny k}} (1+z)^2 + \Omega_{\mbox{\tiny
$\Lambda$}} \right \}^{1/2} \; .
\end{equation}
In equation \ref{eq_vmax}, $\Delta \Omega$ is the solid
angle covered by the LALA survey, and \fplya\ is the
\lya\ line flux for the source in question if it were
located at redshift $z'$. The lower limit of
integration $z_1$ is set by the lowest wavelength at
which \lya\ could be detected by our narrow-band
filters, corresponding to $z \approx 4.37$. The upper
limit of integration $z_2$ is set in one of two ways.
If the \lya\ luminosity for a source is bright enough
that the line remains above the survey flux limit out
to the highest redshift accessible by our filter set,
then $z_2$ is simply equal to the upper redshift limit
for the survey, $z \approx 4.57$. For fainter sources,
$z_2$ is taken to be the redshift at which the \lya\
flux falls below the survey flux limit; in this case,
$4.37 < z_2 < 4.57$.

Having computed \vmax\ for each galaxy, we may compute
the differential \lya\ luminosity function $\Phi(L)$,
the number density of galaxies per logarithmic interval
in \lya\ luminosity.  In a given luminosity bin of
width $\Delta \log L$ centered on $L_i$, this is
given by
\begin{equation}
\Phi (L_i) = \frac{1}{\Delta \log L} \sum_j
\frac{1}{\vmax_{,j}} \; .
\end{equation}
Here, the index $i$ denotes the luminosity bin and $j$
denotes the galaxies within the bin, where the
galaxies summed in a given bin are selected by their
\lya\ luminosities according to
\begin{equation}
| \log L_j - \log L_i | < \frac{\Delta \log L}{2} \; .
\end{equation}
Finally, the uncertainty in the luminosity function may
be estimated with
\begin{equation}
\sigma [ \Phi (L_i)] = \frac{1}{\Delta \log L} \left [
\sum_j \left ( \frac{1}{\vmax_{,j}} \right )^2 \right
]^{1/2} \; .
\label{variance}
\end{equation}

\begin{figure}[!t]
\centering
\epsscale{1.0}
\plotone{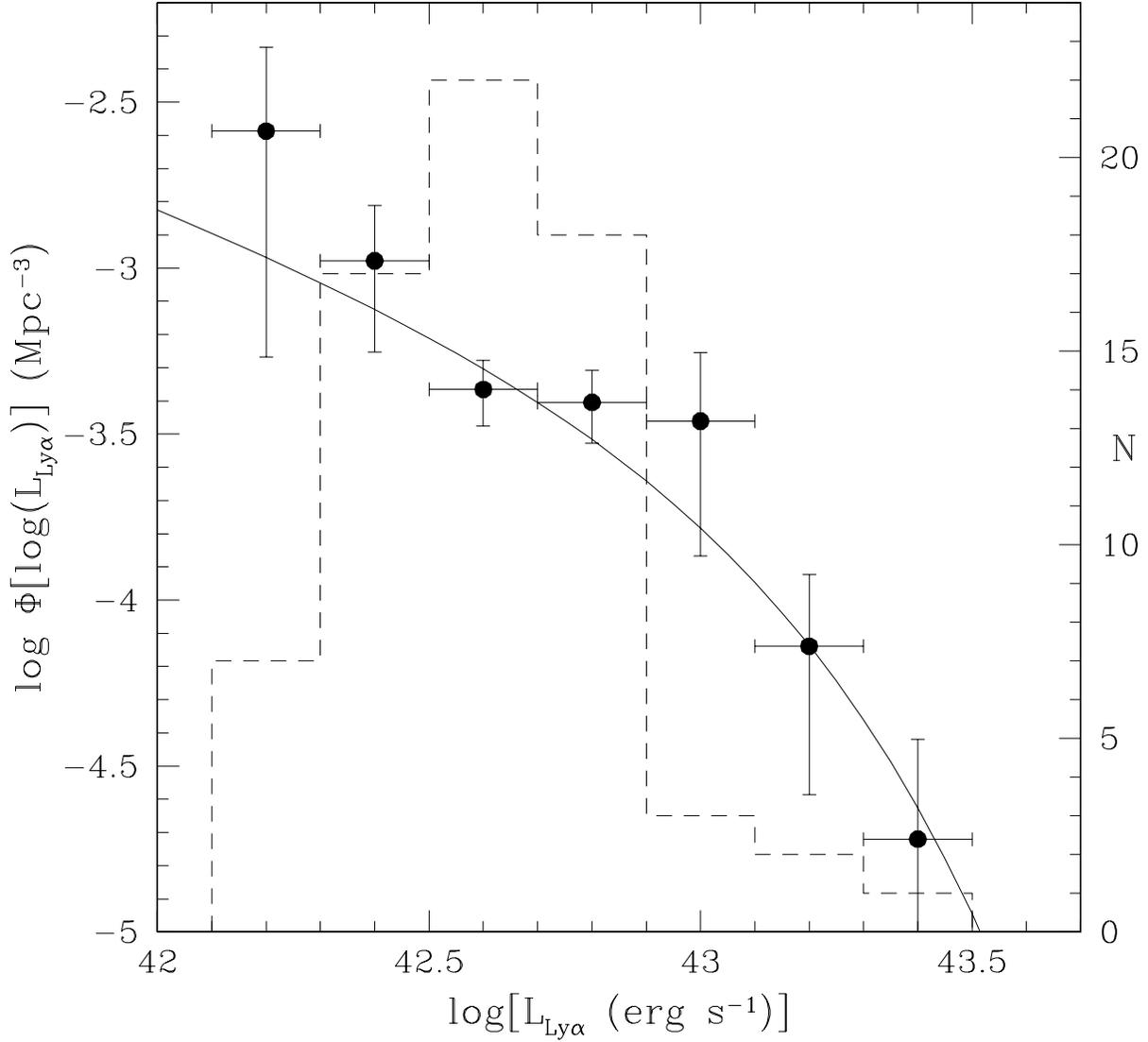}
\caption{Differential \lya\ luminosity function for our
$z=4.5$ sample, computed using the
$V / \vmax$ method.  
The sample includes both the Keck/DEIMOS data presented in 
this paper and the Keck/LRIS data presented in Paper~I. 
The error bars are the 1$\sigma$ 
statistical uncertainties given by the root
variance shown in equation \ref{variance}.
The background histogram (dashed) gives the number of individual
sources contributing to each luminosity bin.
The solid curve shows the best fitting Schechter function
model, with $L^*=(10.9 \pm 3.3) \times 10^{42}$ erg s$^{-1}$, 
$\Phi^* = (1.7 \pm 0.2) \times 10^{-4}$ Mpc$^-3$, and a fixed
faint-end slope $\alpha = -1.6$.}
\label{difflumfuncfig}
\end{figure}

In Figure~\ref{difflumfuncfig}, we present the resulting 
differential \lya\ luminosity function at $z \approx 4.5$. 
We also fit the data with a Schechter function. If $\Phi(L)
\, dL$ is the comoving number density of galaxies with
luminosities in the range $(L,L+dL)$, then the corresponding
Schechter function is
\begin{equation}
\Phi(L) \, dL = \frac{\Phi^*}{L^*} \left (
\frac{L}{L^*} \right )^\alpha \exp \left ( -
\frac{L}{L^*} \right ) \, dL \; ,
\end{equation}
where $\Phi^*$ is the normalization, $L^*$ is the
characteristic break luminosity, and $\alpha$ sets the
slope at the faint end. This is related to the number
density of galaxies in {\it logarithmic} intervals by
\begin{eqnarray}
\Phi(L) \, d(\log L) = \left ( \frac{L}{\log_{10} e}
\right ) \left ( \frac{\Phi^*}{L^*} \right ) \left (
\frac{L}{L^*} \right )^\alpha  \nonumber \exp \left ( -
\frac{L}{L^*} \right ) \, d(\log L) \; ,
\end{eqnarray}
and it is this function which we fit to our data. As in
\citet{vanbreukelen05}, because the binned data points
are few, we choose to fix $\alpha = -1.6$ so as to fit
with only two free parameters, $\Phi^*$ and $L^*$. This
choice fits well with the luminosity distribution of
both LBGs and \lya-emitters at $z \approx 3$
\citep{steidel99,steidel00}.
We find best-fit luminosity function paramters 
$L^* = (10.9  \pm 3.3) \times 10^{42} $erg s$^{-1}$
and $\Phi^* = (1.7 \pm 0.2) \times 10^{-4}$ Mpc$^-3$
(or equivalently, $\log(L^*) = 43.04 \pm 0.14$ and
$\log(\Phi^*) = -3.77 \pm 0.05$). 
The error bars on $L^*$ and $\Phi^*$ are the $1 \sigma$
formal errors computed from the covariance matrix in the nonlinear
least-squares fit, scaled by the measured value of $\chi^2$.  That is,
$\delta L^* = \sigma_{L^*} \sqrt{\chi^2 / n_{\mbox{\tiny DOF}}}$, and
similarly for $\delta \Phi^*$ \citep{num_recs}.

Our $z\approx 4.5$ sample provides one of the best measured
\lya\ luminosity functions to date.  We can study redshift
evolution of the \lya\ luminosity function by comparing to 
results from the literature.  Recognizing that
the uncertainties in $L^*$ and $\Phi^*$ are strongly correlated,
we examine not only the individual parameters but also the 
product $L^* \Phi^*$, which is proportional to \lya\ luminosity 
density, and which generally has smaller uncertainties than
the individual parameters.  For our sample, $\log(L^*\Phi^*) =
39.27$.

At lower redshift, there is a $z\approx 3.1$ LF by \citep{gronwall07},
who fit all three parameters.  They find $\alpha =
-1.49^{+0.45}_{-0.34}$, $\log(L^*) = 42.64^{+0.26}_{-0.15}$, and
$\log(\Phi^*) \approx -2.89 \pm 0.04$,
whence $\log(L^* \Phi^*) \approx 39.75$.
At the high redshift end, we compare to LFs at $z=6.5$ by
\citet{malhotra04} and \citet{kashikawa06}, and at $z=5.7$
by \citet{malhotra04} and \citet{shimasaku06}, all derived by
fixing the faint end slope $\alpha = -1.5$ and fitting 
$L^*$ and $\Phi^*$.  At $z=6.5$, the LFs are similar to our $z=4.5$ 
result: \citet{malhotra04} find $\log(L^*) = 42.6$,
$\log(\Phi^*) = -3.3$, and $\log(L^*\Phi^*) = 39.3$, 
while \citet{kashikawa06} find (for their
combined spectroscopic + photometric sample)  
$\log(L^*) = 42.6$, $\log(\Phi^*) = -2.88$, and $\log(L^*\Phi^*) = 39.72$.
At $z=5.7$, \citet{malhotra04} find $\log(L^*) = 43.0$, 
$\log(\Phi^*) = -4.0$, and $\log(L^*\Phi^*) =39.0$, 
while \citet{shimasaku06} find
$\log(L^*) = 42.9\pm{0.14}$, $\log(\Phi^*) = -3.2 \pm 0.17$, and
$\log(L^*\Phi^*) = 39.7$.  
The obvious differences between the LFs at each redshift may be caused
by any combination of (a) simple uncertainty in deriving the LF from
modest sized samples; (b) field-to-field variations in \lya\ galaxy
density; or (c) differences in the methods used to derive Schechter
function parameters, and in part to local variations in \lya\ galaxy
density.  The \citet{kashikawa06} and \citet{shimasaku06} LFs are
derived from larger total samples, but from a single survey field,
while the \citet{malhotra04} LFs are based on a combination of several
older, smaller samples from a few widely separated fields.
Regardless, if we take the difference between these various
$z\sim 6$ LFs as an empirical indication of total present uncertainties,
the $z\approx 4.5$ LF derived in the present paper supports a 
roughly constant \lya\ luminosity density over the range
$z = 4.5 \pm 1.5$.

\subsection{Comparison to LBGs}

It is interesting to compare this result to the
evolution of the rest-UV luminosity density and cosmic
star formation rate density (SFRD) derived from LBGs
over the same redshift range.
Estimates of the $z\approx 6$ SFRD based on the Great
Observatories Origins Deep Survey/Advanced Camera for
Surveys \citep[GOODS/ACS;][]{giavalisco04} and the 
Hubble Ultra Deep Field \citep{beckwith06} show a factor of 1.5--6 drop 
between $z \sim 3$ and $z \sim 6$
\citep[]{bunker04, bouwens06, giavalisco04sfr}.

What does this mean for the \lya-selected
galaxies? \citet{steidel00} report that \lya-selection
with an equivalent width criterion typical of
narrow-band surveys would return 20\% -- 25\% of their
$z \approx 3$ LBGs.  If \lya-emitters are merely a
subset of the LBG population which happen to have been
detected during a stage of strong \lya\ production,
than we would expect the \lya\ luminosity density
to decline beyond $z\sim 3$, in step with the
global star formation rate density.
Integrating the luminosity functions discussed in 
section~\ref{difflumfunc} shows no compelling evidence 
for such a decline.  Though a modest decline cannot be
firmly ruled out, we may nontheless speculate that the
\lya-emitters as a population are evolving differently
from the LBGs. 

The bolometric luminosities of \lya\ galaxies are typically
lower than Lyman break galaxies, and provide a
hint that they are less massive.  Detailed spectral energy
distribution fitting \citep{gawiser06, finkelstein07,
pirzkal07} bears out this preliminary inference, showing 
typical masses of $\sim 10^8 M_\odot$ and ages $\sim 10^7$
to $10^8$ years.  The correlation strengths of \lya\ galaxies
and LBGs are similar \citep{ouchi03, kovac07},
indicating similar masses of haloes.
From the expected halo mass one can predict volume number density of
\lya\ emitters. Comparing the expected and observed number densities 
implies a duty cycle of \lya-emission in the range 6\% -- 50\% \citep{kovac07}.
A similar duty cycle, 7.5\% - 15\%, is inferred from stellar population 
modelling of the photomeric sample \citep{malhotra02}.

\subsection{Implications for Reionization}

The spectroscopic observations of the $z > 6$ quasars  
yielded the first detections of the long-awaited
Gunn-Peterson trough, implying at least the end of reionization
at $z \approx 6$ \citep{becker01, djorgovski01,
fan02}. Subsequently, the {\it Wilkinson Microwave
Anisotropy Probe} (WMAP) identified a large amplitude
signal in the temperature-polarization maps of the   
cosmic microwave background \citep{spergel03, page06}
indicating a large optical depth to Thomson scattering
and favoring reionization instead at $z \approx 11$.
The WMAP results are not necessarily inconsistent with
those of the quasar Gunn-Peterson troughs. 
Only a small neutral fraction ($x_{\mbox{\tiny HI}}^{\mbox{\tiny IGM}} \sim
0.001$) is required to produce the Gunn-Peterson
effect, so one plausible scenario is that reionization
may have been an extended event, beginning early but not
completing until $z \approx 6$.  Alternatively, a   
variety of theoretical models now suggest that 
reionization occurred twice, first at $z \approx 20$
with the onset of zero-metallicity Population III   
stars, and then again by massive Population II stars  
formed after a partial recombination
\citep[e.g.,][]{cen03, haiman03, somerville03}.

High-redshift \lya-emitting galaxies offer another  
perspective on this issue, as the visibility of   
\lya\ emission should be a sensitive function of the
IGM neutral fraction \citep[e.g.,][]{haiman99, santos04}.
\citet{malhotra04} and \citet{stern05} present first
attempts to exploit this fact by comparing luminosity
functions of \lya-emitters at $z \sim 5.7$ and $z \sim
6.6$. They find no measurable evolution between
these epochs, from which they infer that the IGM
remains largely reionized from the local universe out
to $z \approx 6.5$ \citep[but see][]{haiman05}.
\citet{kashikawa06}, applying the same test, find
possible evidence for observed \lya\ LF differences between
$z=5.7$ and $6.5$ at the factor of 2 level.  They suggest
neutral gas at $z\approx 6.5$ as the explanation, though
\citet{dijkstra07} argue that the observations could equally
well be explained by the ongoing growth of cosmic structure
from $z=6.5$ to $z=5.7$.

By using the \lya\ galaxy sample from \citet{taniguchi05}, 
\citet{malhotra06} showed that at least 30\% of the IGM by volume 
is ionized at $z\approx 6.5$.
This is corroborated by dark gap statistics in GP troughs 
\citep{fan06}. All the \lya\ tests of reionization assume 
that there is no intrinsic evolution in the \lya\ luminosity
functions between $z=5.7$ and 6.5. In this paper we show that
there is little evolution in \lya\ luminosity function 
from $z=6.6$ to $z=3.1$, thus strengthening the conclusion that 
the IGM is not substantially neutral at $z=6.5$.

Significantly, the related question of {\it what} is
responsible for reionization remains at large.  It has
long been recognized that AGN at early epochs are
insufficient, owing to their rapid decline in space
density at high redshift \citep[e.g.,][]{madau99,
barger03}.  Based on their analysis of the UDF,
\citet{bunker04} conclude that the cosmic SFR in 
directly observed $z \approx 6$ LBGs was roughly five times too low to
reionize the Universe. \citet[]{yan04} and \citet[]{bouwens06}
argue that the ionizing photon budget is sufficient provided
one accounts for sample incompleteness using a sufficiently
steep slope at the faint end of the luminosity function.
\citet{malhotra05} argue that the ionizing flux density may
be very inhomogeneous due to large scale structure, as seen 
in galaxies in the Hubble Ultra Deep Field, and that
the directly observed galaxies at $z\approx 6$ do produce 
sufficient photons for reionization in overdense regions.
 
We estimate that the contribution to the cosmic SFR from
\lya-emitters at this epoch is lower than that of the
LBGs ($\rho_{\mbox{\tiny SFR}} (\lya) \approx 0.003$
$M_{\sun}$ yr$^{-1}$ Mpc$^{-3}$, as compared to
$\rho_{\mbox{\tiny SFR}} (\mbox{LBG}) \approx 0.005$
$M_{\sun}$ yr$^{-1}$ Mpc$^{-3}$) when integrated over
the same limits. Consequently, though high-redshift
\lya-emitters are proving to be a useful probe of the
history of reionization, they are evidently not its
cause. While extinction corrections could occasionally be
large \citep{chary05} and could modify this
conclusion, most well studied \lya\ galaxies have very modest
extinction \citep{finkelstein07, pirzkal07}.

When we compare the luminosity function of \lya-emitters at $z
\approx 4.5$ to luminosity functions for similarly assembled samples
spanning $3.1 < z < 6.6$, we find no evidence for evolution over these
epochs.  This result bolsters the conclusion by \citet{malhotra04}
and \citet{stern05} that the IGM remains largely reionized from the
local universe out to $z \approx 6.5$.  However, it is somewhat at
odds with the factor of 1.5--6 drop in the cosmic star formation rate
density measured by \citep[]{bunker04, bouwens06, giavalisco04}
between $z \sim 3$ and $z \sim 6$ in Lyman-break galaxies selected in
the exceptional imaging of the UDF. It seems that these two populations--- 
\lya\ emitters and Lyman Break Galaxies--- follow different evolutionary
histories.  The disentanglement of this issue will
likely rely on extensive followup observations of large samples, 
so that we can study the continuum and absorption lines of many 
\lya\ galaxies, and conversely the \lya\ properties of the break-selected 
galaxies.


\acknowledgements

This work benefited greatly from conversations with M. 
Cooper, S. McCarthy, T. Robishaw, and J. Simon,
as well as from the careful commentary of the anonymous
referee.  
In addition, we are humbly indebted to the expert staff of
W. M. Keck Observatory for their assistance in   
obtaining the data herein. It is a pleasure to thank P.
Amico, J. Lyke, and especially G.  Wirth for their invaluable     
assistance during observing runs. We thank F. Valdes for 
writing the ``deitab'' package, which aids in DEIMOS data
processing.  Finally, we wish to
acknowledge the significant cultural role that the
summit of Mauna Kea plays within the indigenous
Hawaiian community; we are fortunate to have the
opportunity to conduct observations from this mountain.
This material is based upon work supported by AURA through the 
National Science Foundation under AURA Cooperative Agreement 
AST 0132798 as amended. 
The work of D. S. was carried out at the Jet Propulsion
Laboratory, California Institute of Technology, under  
contract with NASA. A. D. and B. J. acknowledge support
from NOAO, which is operated by the Association of
Universities for Research in Astronomy, Inc., under
cooperative agreement with the National Science
Foundation (NSF). H. S. gratefully acknowledges NSF
grant AST 95-28536 and its successors for supporting much of the research
presented herein. This work made use of NASA's
Astrophysics Data System Abstract Service.


\eject


\begin{deluxetable}{llcllcll}
\tablewidth{0pt}
\tabletypesize{\scriptsize}
\tablecolumns{8}
\tablecaption{Spectroscopic Properties}
\tablehead{\colhead{}                                   &
\colhead{}                                              &
\colhead{\lya\ Flux\tablenotemark{b}}                   &
\colhead{\wrst\tablenotemark{c}}                        &
\colhead{FWHM\tablenotemark{d}}                         &
\colhead{$\Delta v$\tablenotemark{e}}                   &
\colhead{Continuum ($\mu$Jy)\tablenotemark{f}}          &
\colhead{Continuum ($\mu$Jy)\tablenotemark{f}}          \\
\colhead{Target}                                        &
\colhead{$z$\tablenotemark{a}}                          &
\colhead{($10^{-17}$ erg cm$^{-2}$ s$^{-1}$)}           &
\colhead{(\AA)}                                         &
\colhead{(\AA)}                                         &
\colhead{(km s$^{-1}$)}                                 &
\colhead{Blue Side}                                     &
\colhead{Red Side}}
\startdata
J020418.2$-$050748 & 4.449 & 2.55 $\pm$ 0.87 & $>86$\tablenotemark{g} & 6.8 $\pm$ 1.7 & 230 & -0.040 $\pm$ 0.026 & -0.018 $\pm$ 0.049 \\ 
J020423.2$-$050647 & 4.449 & 3.25 $\pm$ 1.07 & $>108$\tablenotemark{g} & 5.7 $\pm$ 1.0 & 160 & -0.003 $\pm$ 0.026 & 0.014 $\pm$ 0.034 \\ 
J020425.5$-$045610 & 4.461 & 3.72 $\pm$ 1.22 & $379^{+2092}_{-187}$ & 7.2 $\pm$ 1.0 & 260 & 0.002 $\pm$ 0.025 & 0.026 $\pm$ 0.031 \\ 
J020425.7$-$045810 & 4.387 & 1.98 $\pm$ 0.68 & $>39$\tablenotemark{g} & 4.1 $\pm$ 0.6 & $<$ 200\tablenotemark{h} & -0.035 $\pm$ 0.052 & 0.021 $\pm$ 0.057 \\ 
J020427.4$-$050045 & 4.390 & 1.47 $\pm$ 0.54 & $>142$\tablenotemark{g} & 5.4 $\pm$ 2.8 & 140 & -0.010 $\pm$ 0.018 & -0.011 $\pm$ 0.019 \\ 
J020428.5$-$045924 & 4.390 & 3.57 $\pm$ 1.27 & $508^{+4493}_{-278}$ & 11.0 $\pm$ 2.8 & 460 & -0.008 $\pm$ 0.032 & 0.019 $\pm$ 0.033 \\ 
J020429.8$-$050251 & 4.460 & 1.39 $\pm$ 0.52 & $>22$\tablenotemark{g} & 7.1 $\pm$ 2.2 & 250 & -0.124 $\pm$ 0.070 & 0.012 $\pm$ 0.077 \\ 
J020432.3$-$045519 & 4.360 & 3.13 $\pm$ 1.04 & $>241$\tablenotemark{g} & 4.2 $\pm$ 1.3 & $<$ 210\tablenotemark{h} & -0.003 $\pm$ 0.023 & -0.009 $\pm$ 0.022 \\ 
J142434.9$+$352833 & 4.423 & 1.13 $\pm$ 0.46 & $26^{+30}_{-12}$ & 6.8 $\pm$ 1.1 & 230 & 0.037 $\pm$ 0.077 & 0.117 $\pm$ 0.073 \\ 
J142436.0$+$352600 & 4.464 & 1.81 $\pm$ 0.72 & $38^{+17}_{-16}$ & 5.0 $\pm$ 0.2 & 100 & 0.016 $\pm$ 0.022 & 0.128 $\pm$ 0.025 \\ 
J142438.4$+$352339 & 4.526 & 3.21 $\pm$ 1.28 & $26^{+11}_{-10}$ & 7.0 $\pm$ 0.5 & 240 & 0.175 $\pm$ 0.036 & 0.336 $\pm$ 0.061 \\ 
J142445.2$+$352920 & 4.509 & 1.21 $\pm$ 0.49 & $9^{+4}_{-3}$ & 9.5 $\pm$ 1.5 & 370 & 0.106 $\pm$ 0.041 & 0.350 $\pm$ 0.049 \\ 
J142445.3$+$352450 & 4.475 & 2.47 $\pm$ 0.99 & $>55$\tablenotemark{g} & 5.6 $\pm$ 0.3 & 150 & -0.056 $\pm$ 0.043 & -0.009 $\pm$ 0.065 \\ 
J142445.4$+$352859 & 4.514 & 0.98 $\pm$ 0.40 & $6^{+2}_{-2}$ & 8.2 $\pm$ 2.0 & 310 & 0.174 $\pm$ 0.040 & 0.447 $\pm$ 0.052 \\ 
J142450.1$+$353000 & 4.507 & 4.32 $\pm$ 1.73 & $83^{+70}_{-36}$ & 8.2 $\pm$ 0.6 & 310 & -0.010 $\pm$ 0.050 & 0.141 $\pm$ 0.063 \\ 
J142452.4$+$352613 & 4.411 & 1.97 $\pm$ 0.79 & $97^{+212}_{-46}$ & 6.5 $\pm$ 0.6 & 210 & 0.051 $\pm$ 0.038 & 0.054 $\pm$ 0.047 \\ 
J142458.6$+$353558 & 4.522 & 2.02 $\pm$ 1.06 & $>24$\tablenotemark{g} & 5.6 $\pm$ 0.5 & 150 & -0.026 $\pm$ 0.078 & -0.006 $\pm$ 0.116 \\ 
J142459.8$+$353927 & 4.482 & 1.98 $\pm$ 1.05 & $>59$\tablenotemark{g} & 5.0 $\pm$ 0.5 & 100 & 0.030 $\pm$ 0.041 & -0.009 $\pm$ 0.049 \\ 
J142501.7$+$353652 & 4.496 & 1.38 $\pm$ 0.78 & $43^{+721}_{-24}$ & 6.4 $\pm$ 2.0 & 200 & 0.073 $\pm$ 0.100 & 0.088 $\pm$ 0.140 \\ 
J142502.8$+$353017 & 4.476 & 0.75 $\pm$ 0.31 & $>20$\tablenotemark{g} & 5.6 $\pm$ 0.9 & 150 & 0.052 $\pm$ 0.034 & 0.016 $\pm$ 0.042 \\ 
J142503.4$+$353222 & 4.489 & 0.66 $\pm$ 0.28 & $21^{+18}_{-9}$ & 4.1 $\pm$ 2.2 & $<$ 200\tablenotemark{h} & 0.113 $\pm$ 0.037 & 0.086 $\pm$ 0.044 \\ 
J142506.4$+$353819 & 4.446 & 8.11 $\pm$ 4.26 & $594^{+4407}_{-336}$ & 7.0 $\pm$ 0.3 & 250 & 0.008 $\pm$ 0.054 & 0.037 $\pm$ 0.065 \\ 
J142508.3$+$353952 & 4.511 & 2.59 $\pm$ 1.36 & $175^{+844}_{-93}$ & 8.6 $\pm$ 0.6 & 330 & -0.019 $\pm$ 0.038 & 0.040 $\pm$ 0.048 \\ 
J142508.7$+$353200 & 4.478 & 2.41 $\pm$ 0.96 & $>75$\tablenotemark{g} & 6.0 $\pm$ 0.4 & 180 & 0.032 $\pm$ 0.040 & -0.010 $\pm$ 0.049 \\ 
J142512.0$+$353913 & 4.451 & 1.13 $\pm$ 0.60 & $>30$\tablenotemark{g} & 4.1 $\pm$ 1.4 & $<$ 200\tablenotemark{h} & 0.063 $\pm$ 0.040 & 0.000 $\pm$ 0.050 \\ 
J142512.7$+$353755 & 4.434 & 2.96 $\pm$ 1.56 & $34^{+19}_{-17}$ & 6.1 $\pm$ 0.6 & 190 & 0.201 $\pm$ 0.040 & 0.235 $\pm$ 0.053 \\ 
J142518.0$+$353415 & 4.408 & 5.37 $\pm$ 2.15 & $39^{+18}_{-15}$ & 8.7 $\pm$ 0.8 & 340 & 0.150 $\pm$ 0.062 & 0.370 $\pm$ 0.072 \\ 
J142522.4$+$353553 & 4.519 & 1.79 $\pm$ 0.72 & $39^{+30}_{-16}$ & 7.4 $\pm$ 0.6 & 260 & 0.011 $\pm$ 0.046 & 0.126 $\pm$ 0.053 \\ 
J142525.9$+$352349 & 4.471 & 3.27 $\pm$ 1.34 & $33^{+16}_{-13}$ & 7.0 $\pm$ 0.7 & 240 & 0.072 $\pm$ 0.048 & 0.267 $\pm$ 0.057 \\ 
J142526.2$+$352531 & 4.464 & 2.76 $\pm$ 1.13 & $85^{+105}_{-39}$ & 6.2 $\pm$ 0.4 & 190 & 0.067 $\pm$ 0.050 & 0.087 $\pm$ 0.054 \\ 
J142531.8$+$352652 & 4.482 & 0.94 $\pm$ 0.40 & $18^{+9}_{-7}$ & 7.4 $\pm$ 1.5 & 270 & 0.034 $\pm$ 0.035 & 0.140 $\pm$ 0.041 \\ 
J142532.9$+$353013 & 4.534 & 5.49 $\pm$ 1.00 & $201^{+75}_{-51}$ & 7.1 $\pm$ 0.3 & 250 & 0.005 $\pm$ 0.015 & 0.075 $\pm$ 0.018 \\ 
J142535.2$+$352743 & 4.449 & 6.23 $\pm$ 2.54 & $159^{+173}_{-72}$ & 6.0 $\pm$ 0.2 & 180 & 0.001 $\pm$ 0.048 & 0.106 $\pm$ 0.057 \\ 
J142539.5$+$353902 & 4.432 & 1.52 $\pm$ 0.67 & $240^{+2182}_{-126}$ & 4.0 $\pm$ 1.6 & $<$ 200\tablenotemark{h} & 0.049 $\pm$ 0.019 & 0.017 $\pm$ 0.022 \\ 
J142541.7$+$353351 & 4.409 & 3.24 $\pm$ 1.34 & $108^{+139}_{-50}$ & 5.4 $\pm$ 0.8 & 130 & -0.042 $\pm$ 0.045 & 0.080 $\pm$ 0.050 \\ 
J142542.0$+$352557 & 4.393 & 1.05 $\pm$ 0.44 & $30^{+28}_{-13}$ & 6.4 $\pm$ 1.4 & 210 & 0.028 $\pm$ 0.033 & 0.092 $\pm$ 0.042 \\ 
J142542.6$+$352626 & 4.450 & 1.49 $\pm$ 0.62 & $19^{+10}_{-7}$ & 7.5 $\pm$ 0.9 & 270 & 0.101 $\pm$ 0.043 & 0.215 $\pm$ 0.059 \\ 
J142544.5$+$354325 & 4.533 & 2.84 $\pm$ 1.20 & $131^{+129}_{-60}$ & 7.4 $\pm$ 1.1 & 260 & 0.002 $\pm$ 0.018 & 0.059 $\pm$ 0.030 \\ 
J142546.8$+$354315 & 4.443 & 0.72 $\pm$ 0.33 & $40^{+34}_{-19}$ & 6.3 $\pm$ 1.5 & 200 & 0.012 $\pm$ 0.016 & 0.049 $\pm$ 0.022 \\ 
J142547.8$+$354200 & 4.539 & 1.11 $\pm$ 0.48 & $>56$\tablenotemark{g} & 4.9 $\pm$ 0.8 & 90 & 0.029 $\pm$ 0.017 & 0.007 $\pm$ 0.024 \\ 
J142548.4$+$352740 & 4.546 & 1.21 $\pm$ 0.50 & $>24$\tablenotemark{g} & 5.4 $\pm$ 0.7 & 130 & -0.058 $\pm$ 0.038 & 0.004 $\pm$ 0.067 \\ 
J142555.4$+$353039 & 4.423 & 10.31 $\pm$ 1.88 & $560^{+467}_{-190}$ & 10.8 $\pm$ 0.6 & 450 & 0.025 $\pm$ 0.015 & 0.049 $\pm$ 0.023 \\ 
J142556.7$+$354234 & 4.425 & 2.26 $\pm$ 0.96 & $189^{+284}_{-89}$ & 4.8 $\pm$ 0.6 & 80 & -0.019 $\pm$ 0.020 & 0.032 $\pm$ 0.022 \\ 
J142556.8$+$354215 & 4.426 & 2.85 $\pm$ 1.20 & $162^{+152}_{-74}$ & 5.8 $\pm$ 0.7 & 170 & 0.021 $\pm$ 0.020 & 0.047 $\pm$ 0.022 \\ 
J142559.8$+$353513 & 4.394 & 1.39 $\pm$ 0.28 & $158^{+189}_{-60}$ & 5.6 $\pm$ 0.6 & 160 & 0.005 $\pm$ 0.013 & 0.023 $\pm$ 0.014 \\ 
J142559.8$+$353748 & 4.420 & 4.95 $\pm$ 0.91 & $55^{+10}_{-10}$ & 7.0 $\pm$ 0.5 & 240 & 0.109 $\pm$ 0.018 & 0.240 $\pm$ 0.019 \\ 
J142601.3$+$353618 & 4.475 & 1.41 $\pm$ 0.27 & $>171$\tablenotemark{g} & 5.0 $\pm$ 0.4 & 100 & -0.013 $\pm$ 0.011 & -0.003 $\pm$ 0.013 \\ 
J142602.0$+$354554 & 4.473 & 1.83 $\pm$ 0.78 & $85^{+70}_{-38}$ & 4.8 $\pm$ 0.7 & 70 & -0.003 $\pm$ 0.019 & 0.058 $\pm$ 0.025 \\ 
J142612.2$+$353541 & 4.418 & 1.90 $\pm$ 0.36 & $>140$\tablenotemark{g} & 6.3 $\pm$ 0.5 & 200 & 0.082 $\pm$ 0.013 & 0.002 $\pm$ 0.017 \\ 
J142624.4$+$353832 & 4.460 & 2.46 $\pm$ 0.46 & $320^{+1182}_{-145}$ & 5.4 $\pm$ 0.5 & 140 & -0.012 $\pm$ 0.017 & 0.021 $\pm$ 0.022 \\ 
J142627.5$+$353717 & 4.488 & 2.11 $\pm$ 0.43 & $42^{+24}_{-14}$ & 6.5 $\pm$ 2.4 & 210 & 0.044 $\pm$ 0.025 & 0.135 $\pm$ 0.051 \\ 
J142628.5$+$353809 & 4.409 & 3.82 $\pm$ 0.71 & $>143$\tablenotemark{g} & 6.4 $\pm$ 0.4 & 210 & -0.038 $\pm$ 0.029 & -0.011 $\pm$ 0.041 \\ 
J142653.5$+$353356 & 4.494 & 2.12 $\pm$ 0.77 & $142^{+115}_{-59}$ & 7.8 $\pm$ 1.8 & 280 & -0.016 $\pm$ 0.015 & 0.040 $\pm$ 0.017 \\ 
J142658.8$+$353144 & 4.495 & 2.16 $\pm$ 0.79 & $31^{+11}_{-10}$ & 6.3 $\pm$ 1.0 & 200 & -0.019 $\pm$ 0.015 & 0.191 $\pm$ 0.018 \\ 
J142706.3$+$353224 & 4.480 & 0.99 $\pm$ 0.38 & $100^{+139}_{-45}$ & 5.0 $\pm$ 1.6 & 100 & 0.021 $\pm$ 0.014 & 0.027 $\pm$ 0.018 \\ 
J142709.1$+$352738 & 4.407 & 1.77 $\pm$ 0.65 & $85^{+50}_{-34}$ & 8.4 $\pm$ 1.4 & 320 & 0.039 $\pm$ 0.014 & 0.056 $\pm$ 0.017 \\ 
J142709.2$+$352409 & 4.520 & 1.62 $\pm$ 0.59 & $>85$\tablenotemark{g} & 5.8 $\pm$ 1.2 & 170 & 0.029 $\pm$ 0.015 & 0.008 $\pm$ 0.022 \\ 
J142709.8$+$352641 & 4.405 & 1.78 $\pm$ 0.66 & $178^{+337}_{-82}$ & 6.0 $\pm$ 0.8 & 180 & 0.023 $\pm$ 0.017 & 0.027 $\pm$ 0.021 \\ 
J142712.2$+$353029 & 4.380 & 2.35 $\pm$ 0.86 & $82^{+45}_{-32}$ & 4.3 $\pm$ 2.4 & $<$ 200\tablenotemark{h} & 0.035 $\pm$ 0.017 & 0.076 $\pm$ 0.021 \\ 
\enddata
\label{table_spec_prop}
\tablenotetext{a}{The redshift was derived from the
wavelength of the peak pixel in the line
profile smoothed with a 3-pixel boxcar average.
We estimate the error in this measurement to be
$\delta_z \approx 0.0005$, based on Monte Carlo simulations
in which we added random noise to each pixel of every
spectrum according to the photon counting statistics,
and then re-measured the redshift in each case.
This measurement may overestimate the true redshift of
the system since the blue wing of the \lya\ emission is
absorbed by foreground neutral hydrogen.}
\tablenotetext{b}{The line flux was determined by
totaling the flux of the pixels that fall within the
line profile.  No attempt was made to model the
emission line or to account for the very minor
contribution of the continuum to the line.  Quoted
uncertainties account for photon counting errors alone,
excluding possible systematic errors. Despite these
caveats, the \lya\ line fluxes measured from the
spectra agree with narrow band imaging 
to $1\sigma$ in all but three cases.}
\tablenotetext{c}{The rest frame equivalent widths were
determined with $\wrst = (F_{\ell} / f_{\lambda,r}) /
(1+z)$, where $F_{\ell}$ is the flux in the emission
line and $f_{\lambda,r}$ is the measured red-side
continuum flux density.  The error bars $\delta
w_{\mbox{\tiny +}}$ and $\delta w_{\mbox{\tiny --}}$
are $1 \sigma$ confidence intervals determined by
integrating over the probability density functions
$P_i(w)$ described in \S~\ref{ew_update}.  The error bars are
symmetric in probability density-space in the sense
that $\int_{w - \delta w_{\mbox{\tiny --}}}^w P_i(w')
\, dw' = \int^{w + \delta w_{\mbox{\tiny +}}}_w P_i(w')
\, dw'= 0.34$.}
\tablenotetext{d}{The FWHM was measured directly from
the emission line by counting the number of pixels in
the unsmoothed spectrum which exceed a flux equal to
half the flux in the peak pixel.  No attempt was made
to account for the minor contribution of the continuum
to the height of the peak pixel.
The error bars
were determined with Monte Carlo simulations in which
we modeled each emission line with the truncated
Gaussian profile described in \citet{hu04} and
\citet{rhoads04}, added random noise in each pixel
according to the photon counting errors, and then measured
the widths $\sigma(\mbox{FWHM})$ of
the resulting distribution of FWHM
for the given line.}
\tablenotetext{e}{The velocity width $\Delta v$ was
determined by subtracting in quadrature the effective
instrumental resolution for a point source, and is
therefore an upper limit, as the target may have
angular size comparable to the $\simlt 1\arcsec$ seeing
of these data. Where the emission line is unresolved,
the velocity width is an upper limit set by the
effective width of the resolution element itself.}
\tablenotetext{f}{Red and blue side continuum
measurements are variance--weighted averages made in
1200 \AA\ wide windows beginning 30 \AA\ from the
wavelength of the peak pixel in the emission line.  
We employed a 10--iteration, $2 \sigma$ clipping
algorithm to reduce the effect of spurious
outliers occurring at long wavelength, where the
sky noise is large.  In some cases,
a small correction factor was subtracted from the
variance--weighted averages based on the detection of
residual signal remaining in extractions of
source--free, sky--subtracted regions of the
two--dimensional spectra (see text,
\S~\ref{obs_spectroscopy}). Quoted uncertainties
account for photon counting errors in the source
extractions added in quadrature to the photon counting
errors derived in the blank--sky extractions.}
\tablenotetext{g}{$2 \sigma$ lower limit.  The
measurement of the red--side continuum for this source
is formally consistent with no observable flux.  The
equivalent width limit was then set by using a $2
\sigma$ upper limit to $f_{\lambda,r}$ in the
expression given in footnote (c).}
\tablenotetext{h}{This line is unresolved.}
\end{deluxetable}

\end{document}